%% file: GJu_MOVPE_apparatus_v6revised1.tex
\documentclass[aip,rsi,reprint]{revtex4-1}   
\pdfoutput=1
\usepackage{graphicx}
\usepackage{dcolumn}
\usepackage{bm}
\usepackage{amsmath, amssymb, euscript}
\usepackage[colorlinks,citecolor=red,urlcolor=blue]{hyperref}

\graphicspath{{figures_pdf/v6revised/}}    

\begin{document}

\title{An instrument for \textit{in situ} coherent x-ray studies of metal-organic vapor phase epitaxy of III-nitrides}

\author{Guangxu Ju}
\email[Author to whom correspondence should be addressed. Electronic mail:]{gju@anl.gov}

\author{Matthew J. Highland}
\affiliation{Materials Science Division, Argonne National Laboratory, Argonne, IL 60439, USA}

\author{Angel Yanguas-Gil}
\affiliation{Energy Systems Division, Argonne National Laboratory, Argonne, IL 60439, USA}

\author{Carol Thompson}
\affiliation{Department of Physics, Northern Illinois University, DeKalb IL 60115, USA}

\author{Jeffrey A. Eastman}
\affiliation{Materials Science Division, Argonne National Laboratory, Argonne, IL 60439, USA}

\author{Hua Zhou}
\affiliation{X-ray Science Division, Argonne National Laboratory, Argonne, IL 60439, USA}

\author{Sean M. Brennan}
\affiliation{Fairview Associates, Jackson, WY 83002, USA}

\author{G. Brian Stephenson}
\email{gbs@anl.gov}

\author{Paul H. Fuoss}
\affiliation{Materials Science Division, Argonne National Laboratory, Argonne, IL 60439, USA}

\date{2017 February 27}

\begin{abstract}
We describe an instrument that exploits the ongoing revolution in synchrotron sources, optics, and detectors to enable \textit{in situ} studies of metal-organic vapor phase epitaxy (MOVPE) growth of III-nitride materials using coherent x-ray methods. The system includes high-resolution positioning of the sample and detector including full rotations, an x-ray transparent chamber wall for incident and diffracted beam access over a wide angular range, and minimal thermal sample motion, giving the sub-micron positional stability and reproducibility needed for coherent x-ray studies. The instrument enables surface x-ray photon correlation spectroscopy, microbeam diffraction, and coherent diffraction imaging of atomic-scale surface and film structure and dynamics during growth, to provide fundamental understanding of MOVPE processes.
\end{abstract}

\maketitle

\section{Introduction}

\textit{In situ} x-ray scattering is a powerful tool for investigating the mechanisms of epitaxial film growth at an atomic scale. By revealing the development of atomic-scale surface morphology, strain, and defects in real time during growth, it provides understanding of the impact of growth conditions on film and interface structures, and thus optical or electronic properties. Examples of film growth processes for III-nitride semiconductors studied to date using \textit{in situ} x-ray techniques include
metal-organic vapor phase epitaxy (MOVPE),
\cite{1999_Stephenson_MRSBull24_21,
2006_Jiang_APL89_161915,
2014_Perret_APL105_051602,
2011_takeda_2011x,
iida2013analysis,
2014_Ju_JCrystGrowth407_68,
2014_ju_JAP_insitu}
molecular beam epitaxy (MBE),
\cite{1998_Headrick_PRB58_4818,
1999_Woll_PRL83_4349,
2016_Sasaki_JJAP55_05FB05}
and sputtering.
\cite{2001_Kang_JMaterRes16_1814} 

Research programs using standard laboratory x-ray sources \cite{2011_takeda_2011x,
iida2013analysis,
2014_Ju_JCrystGrowth407_68,
2014_ju_JAP_insitu,
2001_Kang_JMaterRes16_1814} 
for \textit{in situ} studies typically monitor specular reflectivity or diffraction in the Bragg geometry, which gives sufficient intensity for real-time measurements. The chambers and diffractometers used are primarily designed for symmetric reflection conditions, and they allow observation of the evolution of film thickness, surface or interface roughness of a heteroepitaxial structure, and superlattice uniformity. Composition or film relaxation can be inferred from the evolution of Bragg peak positions and widths. 

\textit{In situ} studies at a synchrotron 
\cite{1999_Stephenson_MRSBull24_21,
2006_Jiang_APL89_161915,
2014_Perret_APL105_051602,
1998_Headrick_PRB58_4818,
1999_Woll_PRL83_4349,
2016_Sasaki_JJAP55_05FB05} 
can employ the higher x-ray brilliance for surface-sensitive techniques such as grazing incidence x-ray scattering,\cite{1990_Fuoss_AnnRevMatSci20_365} as well as fast-scan reciprocal space mapping and observation of weak features such as crystal truncation rods (CTRs), fractional-order peaks, diffuse scattering and diffraction from ultrathin films. For this purpose, various research groups have designed chambers with larger windows mounted on four- or six-circle diffractometers that provide access to asymmetric diffraction conditions and a wide range of reciprocal space. These studies are sensitive to atomic-scale surface reconstructions, step morphology, island size distributions, domain structures, and their evolution during growth. Observation of both in-plane and out-of-plane peak positions and widths allows determination of both composition and film relaxation. 

\begin{figure} [hb!]
	\includegraphics[width=\columnwidth]{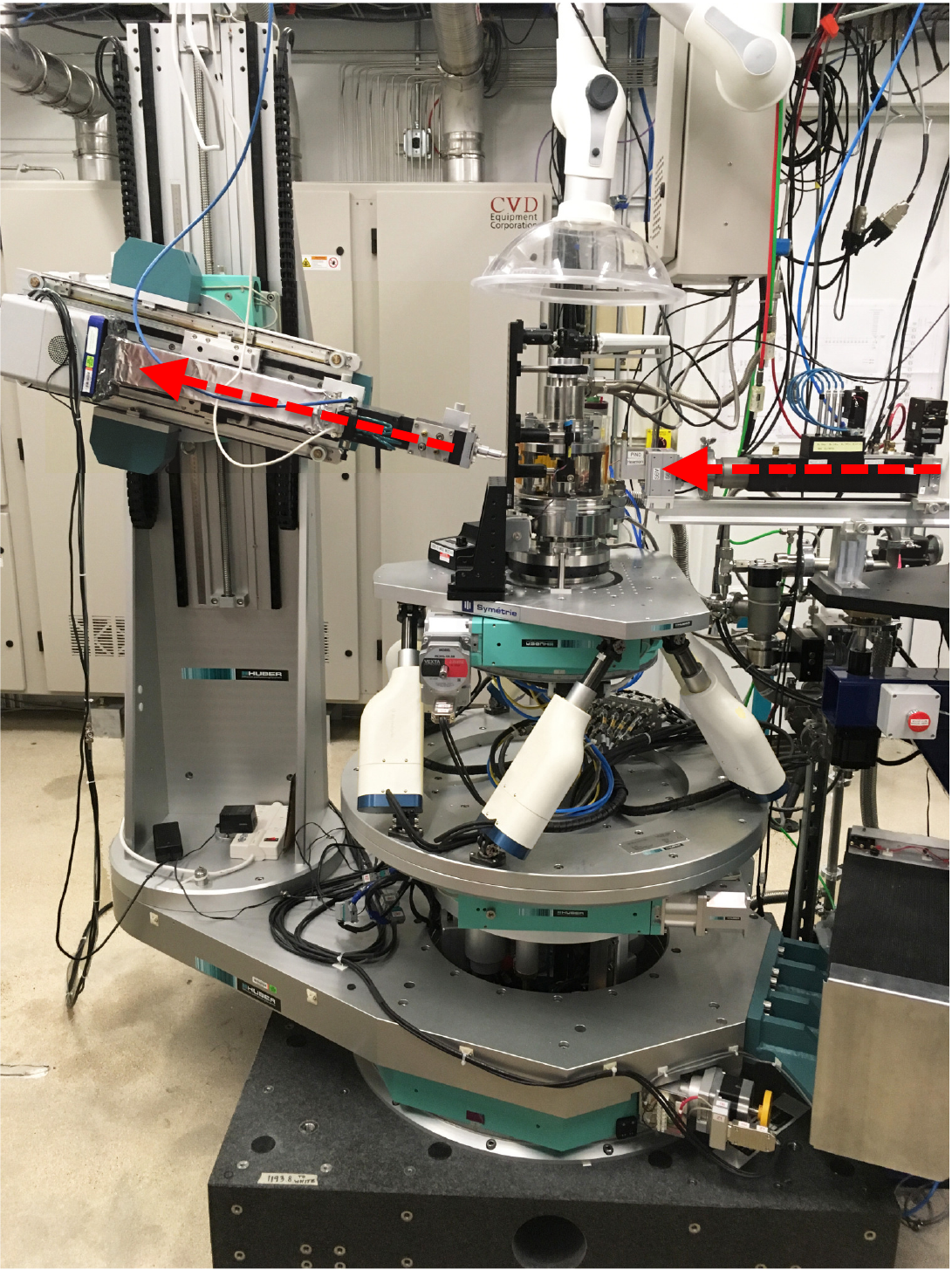}
	\caption{\label{figure1} Photograph of the instrument installed at Advanced Photon Source station 12ID-D. Red dashed arrows show trajectory of x-rays entering through incident beam optics, scattering from the sample in the MOVPE chamber, and being imaged by an area detector mounted on the detector positioner. 
	}
\end{figure}

The system described in this paper, shown in Fig.~\ref{figure1}, was designed to take advantage of an emerging revolutionary development in synchrotron x-ray methods -- the use of coherent x-ray beams. Coherent beam methods allow study of the spatial arrangement of defects and distortions in a crystalline film, including atomic-scale features on its surface, by observing the complex diffraction patterns they produce, known as ``speckle''. These techniques are sensitive to the exact arrangement of nanoscale structures, rather than just spatially averaged quantities. X-ray photon correlation spectroscopy (XPCS) reveals the atomic-scale dynamics by analysis of the intensity fluctuations in the speckle pattern.\cite{2009_stephenson_naturematerials,2014_Shpyrko_JSynchRad21_1057} The dynamics of surface steps and islands revealed by XPCS\cite{pierce2009surface} will shed new light on epitaxial growth mechanisms. Inversion of the speckle pattern to obtain an image of the structure, known as coherent diffraction imaging (CDI),\cite{2013_Abbey_JOM65_1183} will provide a powerful new tool for \textit{in situ} studies of phenomena such as step motion,\cite{2015_Zhu_APL106_101604} 
island nucleation and growth, dislocation and grain boundary dynamics, and strain relaxation.\cite{hruszkewycz2012quantitative}

\begin{figure}
    \centering
    \includegraphics[width=\columnwidth]{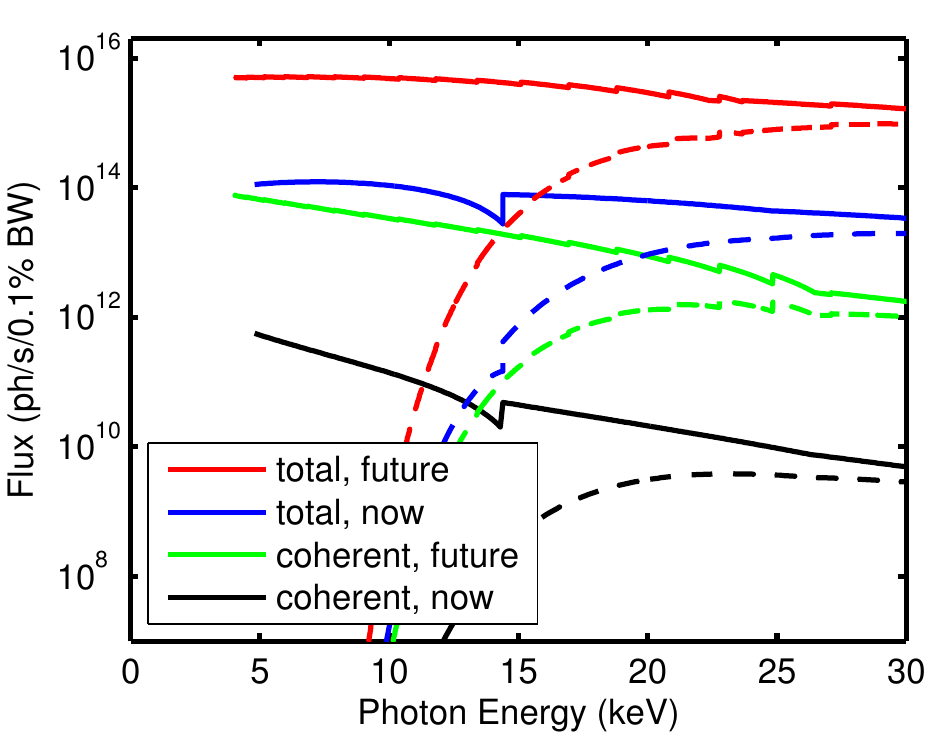}
    \caption{\label{figure2} The total and coherent flux as a function of photon energy from the current undulator source at beamline 12ID-D of APS (3.0 cm period, 2.0 m length)\protect{\cite{Dejus_private}}, as well as the expected values from undulator sources provided by the future APS Upgrade.\protect{\cite{2015_APS_Early_Science}} Dashed curves show fluxes transmitted through both 2-mm-thick fused quartz walls of the MOVPE chamber.}
\end{figure}

The rapid development of coherent x-ray techniques is strongly driven by orders-of-magnitude increases in available coherent x-ray flux. Figure~\ref{figure2} shows the total flux and coherent flux as a function of photon energy from the current undulator source at beamline 12ID-D of the Advanced Photon Source (APS), as well as the much higher values that will be provided by the APS Upgrade.\cite{2015_APS_Early_Science}
The values for the flux transmitted through our MOVPE chamber walls (4 mm total of fused quartz) are indicated by the dashed curves. The optimum photon energy to maximize transmitted flux is about 30 keV for total flux and 23 keV for coherent flux. One can see that the coherent flux available from the APS Upgrade will approach the total flux now available, so all of the surface x-ray scattering techniques currently in use will be feasible with coherent x-ray beams. 

Anticipating the advent of these capabilities, we have developed a next-generation instrument for \textit{in situ} synchrotron x-ray studies of MOVPE. The new instrument is designed to provide the high accuracy and stability required for coherent x-ray scattering techniques during studies under MOVPE conditions. Here we describe the MOVPE chamber, diffractometer, and growth system, and its characterization and initial use at beamline 12ID-D of the APS. 

\section{\label{sec:chamber} System Design}
\subsection{Growth chamber}
\begin{figure*}
	\includegraphics[width=2\columnwidth]{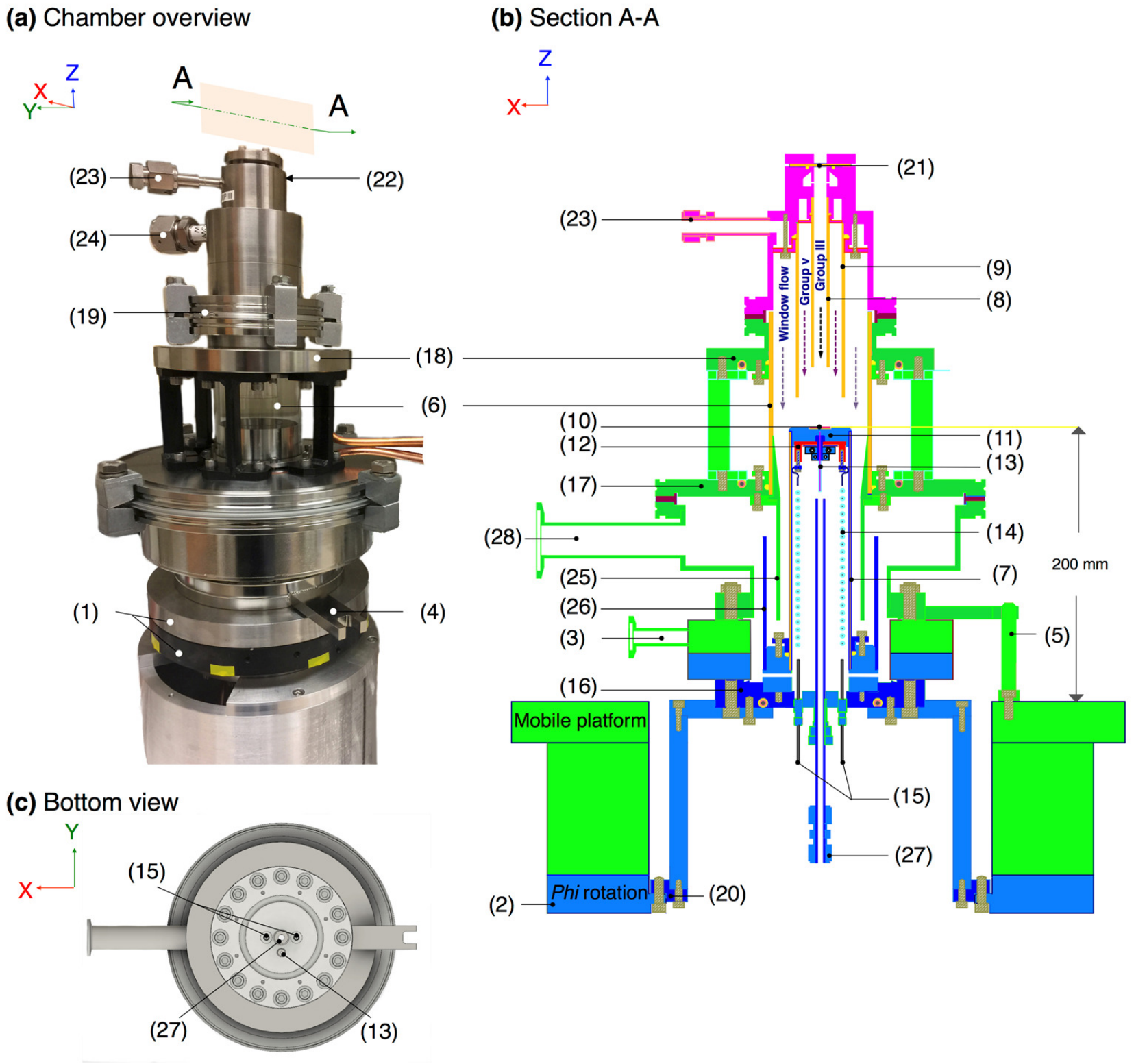}
	\caption{\label{figure3} Main features of the MOVPE chamber designed for \textit{in situ} x-ray studies. (a) Photograph of the assembled chamber. (b) Section through the axis of the chamber. (c) View of the bottom flange from below. The three main sections of the chamber (inlet section, window section, and sample heater section) are colored magenta, green, and blue in (b). Numbered items, explained in more detail in the text, are 
	(1) rotary seal,
    (2) \textit{Phi} goniometer circle,
	(3) rotary seal pump out,
	(4) fork,
	(5) pin,
	(6) fused quartz tube for chamber wall,
	(7) fused quartz tube for sample mount,
	(8) input flow tube for group III,
	(9) input flow tube for group V,
	(10) sample,
	(11) molybdenum sample mounting block or susceptor,
	(12) heater,
	(13) thermocouple (type-K),
	(14) molybdenum leads with ceramic beads,
	(15) feedthroughs for the heater power,
	(16) bottom flange of the chamber,
	(17) bottom flange of the window,
	(18) top flange of the window, 
	(19) ISO flange,
	(20) bolt circle,
	(21) optical port,
	(22) inlet port for group V flow (hidden),
	(23) inlet port for group III flow,
	(24) inlet port for window curtain flow,
	(25) funnel-shaped section,
	(26) concentric baffles, 
	(27) heater purge inlet flow,
	(28) exhaust port.}
\end{figure*}


The growth chamber must provide the appropriate flow geometry to introduce and mix the precursors at the sample surface, be able to reach the temperatures and pressures needed for MOVPE growth of III-nitrides, as well as be compatible with the surface x-ray scattering geometry and the stability requirements for the coherent experiments. The overall design goals for the growth chamber are summarized here:\\
a) High resolution angular positioning of the sample with full $\pm 180^{\circ}$ rotation about the surface normal;\\
b) X-ray transparent chamber wall providing access for incident and scattered x-ray beams over a wide range of angles;\\
c) Compact design to fit on an x-ray goniometer and minimize precursor and carrier gas usage;\\
d) Chamber materials compatible with 1250 $^\circ$C sample temperature and NH$_3$ environment;\\
e) Minimal thermal drift of sample, and positional stability to the micron / microradian level after thermal equilibration;\\
f) Separate input flow channels for group III and group V with adjustable mixing length in order to avoid parasitic pre-reactions;\cite{nakamura1991novel}
\\
g) Purge flow channels arranged to minimize deposition on the x-ray window and heater;\\
h) Access for changing sample;\\
i) Ability to easily remove the chamber from the goniometer;\\
j) Access for additional optical probes;\\
k) Safety, e.g. chamber and window strength.\\ 

To achieve these goals, a specialized MOVPE chamber design has been developed. Many of the design features are adopted from previous generations of chambers designed for \textit{in situ} studies by members of our group.\cite{1990_Brennan_NuclInstMeth291_86,1999_Stephenson_MRSBull24_21} A diagram of the new chamber is shown in Fig. \ref{figure3}. It employs a stagnation point, vertical-downward-flow geometry with concentric circular cross sections and a horizontal sample surface. Its essential features, labeled in Fig.~\ref{figure3}, are:

A differentially-pumped rotary seal\cite{Thermionics} (1) couples a \textit{Phi} goniometer circle (2) to the chamber, and enables the full $\pm 180^{\circ}$ rotation of the sample and sample heater section for surface x-ray scattering experiments. The rotary seal pump out (3) incorporates a pumped inner stage and a N$_2$ pressurized outer stage to avoid leakage in either direction through the seal. In Fig. \ref{figure3}(b), the chamber sections that rotate together with \textit{Phi} are represented primarily as blue.

The upper chamber sections, consisting of the window section (green) and the inlet section (magenta), are rotationally fixed with respect to the hexapod mobile platform by a fork (4) and pin (5) coupling.  These upper sections do not rotate with \textit{Phi} when the sample/sample heater rotates. The gas inlet, exhaust, and cooling water piping thus require only limited flexibility to follow the tilting and translation of the hexapod mobile platform during scattering experiments.

Fused quartz tubes assembled with o-ring seals are used for the following items: the  x-ray transparent chamber wall (6); the low-thermal-expansion support for the heated sample mount (7); and the input flow tubes for group III (8) and group V (9) channels. The quartz tube forming the x-ray transparent chamber wall has a 70 mm inner diameter and 2 mm wall thickness. This is sufficiently transparent at x-ray photon energies above 20 keV, while being sufficiently strong to have a safe operating pressure range from vacuum to 3.75 atm above ambient (55 psig). Fused quartz x-ray windows are a good choice for coherent x-ray methods at these high photon energies because they do not have internal or surface inhomogeneities that could distort the transmitted wavefront. Input flow tubes of different lengths can be used to adjust the path length of the interdiffusion region. All o-ring seals in the system are made of nitrile polymer to be compatible with NH$_3$.

The sample (10) rests in a 0.25 mm deep, 15 mm $\times$ 15 mm square indentation in the molybdenum sample mounting block (11), which itself sits on the low-thermal-expansion fused quartz support tube (7). A resistive heater (12) made of boron-nitride-encased graphite\cite{Boralectric} is attached to the underside of the sample mounting block. The lower face of the heater is insulated with heat shields made of thin molybdenum sheets. A type-K thermocouple (13) encased in an inconel sheath is embedded in the molybdenum sample mounting block and exits the chamber through an o-ring seal fitting in the bottom flange. Electrical power for the heater is provided by molybdenum leads (14) insulated by ceramic beads that connect to feedthroughs (15) in the bottom flange. The thermocouple and heater power leads are hand-formed into a loose spiral inside the support quartz. The spiral provides a spring tension to hold the sample mounting block firmly in place on the quartz support during operation. During assembly and disassembly, the block can be pulled upward a sufficient distance to access the thermocouple and heater power connections. 

The bottom flange (16) on which the quartz support rests, as well as to the bottom (17) and top (18) flanges of the window, are water cooled. For sample loading and unloading, the chamber top opens at an ISO flange (19) separating the inlet section (magenta) from the window section (green). The assembled chamber can be mounted or dismounted from the \textit{Phi} goniometer using the bolt circle (20) attached to the sample heater section. 

A 10-mm-diameter on-axis port (21) on the inlet section provides a view normal to the sample surface through the group III input tube. Typically an optical lens is mounted and the port may be used for near-normal incidence optical measurements, such as interferometric measurements of film growth or temperature calibration, as described in Section~\ref{sec:char}. Alternatively, an x-ray transparent Be window can be mounted to allow monitoring of x-ray fluorescence, including lower energies that would not penetrate the quartz chamber wall. 

The chamber top inlet section has three inlet-flow ports for the group V (22) and group III (23) flows, and a window curtain flow (24) that minimizes deposition on the x-ray transparent chamber walls. Distribution channels inside the inlet section inject the flows symmetrically around the chamber axis. A funnel-shaped section of the chamber wall (25) below the window and concentric baffles downstream of the sample (26) are designed to maintain a radial flow pattern over the sample. A heater purge inlet flow (27) keeps reactive gases from entering the inside of the heater support. A tube injects this flow just below the heater, and it exits into the main chamber flow near the bottom flange. All flows exit the chamber to the pressure control valve and pump through a relatively large (20 mm diameter) exhaust port (28).

\subsection{X-Ray Detector and Sample Positioning}

X-ray measurements require independent, high resolution positioning of a sample and detector system in a six-dimensional space (three positional coordinates and three angular components). Traditionally, the motions of both the sample and detector have been integrated into a single x-ray diffractometer. However, emerging x-ray techniques have very different requirements for stability, and in particular relative stability, of the sample and detector. For coherent x-ray scattering measurements utilizing an area detector, the stability of the sample with respect to the incident x-ray beam must be better than a fraction of the beam size, e.g. one micron, over the course of the measurement. At the same time, the area detector position must be stable to a fraction of a pixel (e.g. 25 microns) at a long distance from the sample position (typically a meter or greater) so that x-ray speckles can be resolved. These requirements have driven instrument design for coherent x-ray measurements to use separate sample and detector positioners. In addition to stability, advanced coherent diffraction imaging techniques such as ptychography\cite{2015_Zhu_APL106_101604} require that the sample be reproducibly raster-scanned with respect to the incident beam with sub-micron repeatability.

However, in addition to coherent x-ray measurements, a goal of the current instrument design is also to support full-field imaging \cite{2014_Laanait_JSynchRad21_1252} based on new x-ray optics (e.g. compound refractive lenses). For full field imaging, the relative stability of the detector optics and sample position become critical while the position of the sample with respect to the incident beam is much less critical (the opposite of that required for coherent x-ray scattering). In addition, being able to position the imaging optics close to the sample position is required to get appropriate numerical apertures and spatial resolutions.

\begin{figure}[t!]
\includegraphics[width=\columnwidth]{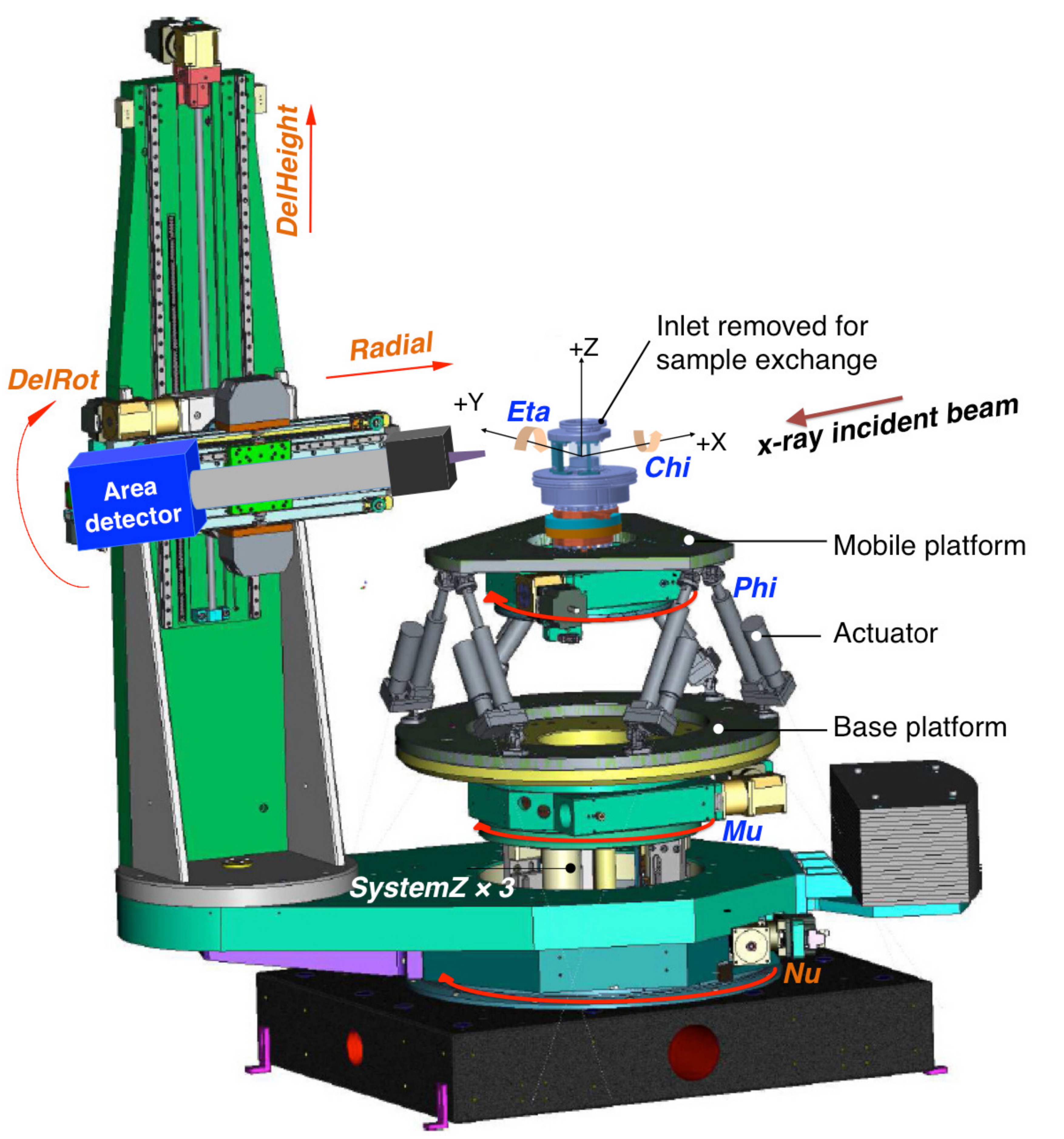} 
\caption{ \label{figure4} A schematic showing the arrangement of the motions provided by the diffractometer system. The sample and its support can independently rotate within the chamber by a full $\pm 180^{\circ}$ in \textit{Phi}. The hexapod mobile platform provides translations of the whole chamber in the  \text{X}, \text{Y}, and \text{Z} directions, and limited rotations in \textit{Eta} and \textit{Chi}. The \textit{Mu} circle provides rotation of the hexapod base platform. The overall height of the combined chamber/hexapod unit can also be adjusted through the \textit{SystemZ} motion. The detector is positioned by separate \textit{Delta} and \textit{Nu} rotations about horizontal and vertical axes, respectively. The \textit{Delta} motion is a compound rotation created by the \textit{DelRot} goniometer and the \textit{DelHeight} translation. The distance between the sample and detector is controlled by the \textit{Radial} translaton. 
}
\end{figure}

To achieve these competing goals, a diffractometer system has been designed with three separate components to position the sample, a near detector, and a far detector. The first two of these are shown in Fig.~\ref{figure4}.

The first component is a high-resolution, high-stability system to move the sample. This system must be capable of moving a relatively large growth chamber with high accuracy. In order to meet future needs, it must also be relatively easy to swap out the growth chamber for a different one (e.g. a different MOVPE chemistry or an entirely different processing technique like sputter deposition). After carefully analyzing requirements, we considered several different motion solutions. The use of complete goniometers to provide angular motions was rejected because prototype designs didn't allow for suitable sample access. The use of arcs was rejected because the necessary stability was difficult and expensive to achieve. After considering several compound motion solutions, it appeared that a hexapod was the most practical solution. However, a hexapod with sufficient total Z motion and rotation about the \textit{Phi} axis was difficult to produce with the required system stiffness.

\begin{table}[htbp]
\caption{Performance of the goniometers\cite{Huber} and angular pseudo-motions\cite{Symmetrie}. Note that the practical range of \textit{Nu} is -10$^{\circ}$ to 100$^{\circ}$ because of collision constraints imposed by MOVPE components, and the range of the combined \textit{Delta} motion is -15$^{\circ}$ to +43$^{\circ}$ due to limits of \textit{DelHeight}.}
\begin{center}
\begin{ruledtabular}
\begin{tabular}{c c c c c}
Motion & Model & Range & Repeatability & Accuracy \\
\hline
\textit{Phi}   & Huber 430 & $\pm$ 180$^{\circ}$              & 0.0006$^{\circ}$  & 0.0028$^{\circ}$ \\
\textit{Chi}   & hexapod   & $\pm$ 5$^{\circ}$                & 0.00006$^{\circ}$ & 0.0003$^{\circ}$ \\
\textit{Eta}   & hexapod   & -5$^{\circ}$ to +24.5$^{\circ}$  & 0.00018$^{\circ}$ & 0.0012$^{\circ}$ \\
\textit{Mu}    & Huber 440 & $\pm$ 180$^{\circ}$              & 0.0006$^{\circ}$  & 0.0028$^{\circ}$ \\
\textit{DelRot} & Huber 420 & $\pm$ 180$^{\circ}$             & 0.0006$^{\circ}$  & 0.0033$^{\circ}$ \\ 
\textit{Nu}    & Huber 480 & $\pm$ 180$^{\circ}$              & 0.0006$^{\circ}$  & 0.0028$^{\circ}$ \\
\end{tabular}
\end{ruledtabular}
\end{center}
\label{table1}
\end{table}

\begin{table}[htbp]
\caption{Performance of the system translations.}
\begin{center}
\begin{ruledtabular}
\begin{tabular}{c c c c}
Motion & Range  & Repeatability & Accuracy \\
\hline
\textit{SampleX}  & $\pm$  10 mm  & 0.23 $\mu$m & 3.31 $\mu$m \\
\textit{SampleY}  & $\pm$  10 mm  & 0.55 $\mu$m & 3.62 $\mu$m \\
\textit{SampleZ}  & $\pm$   5 mm  & 0.07 $\mu$m & 1.18 $\mu$m \\
\textit{SystemZ}  & $\pm$ 125 mm  & 2    $\mu$m & 2    $\mu$m \\
\textit{DelHeight} & -200 +700 mm & 10   $\mu$m & 10   $\mu$m \\
\textit{Radial}   & $\pm$ 250 mm  & 100  $\mu$m & 100  $\mu$m \\
\end{tabular}
\end{ruledtabular}
\end{center}
\label{table2}
\end{table}

Thus, a combination of hexapod and goniometer motions was chosen (see Tables~\ref{table1} and \ref{table2} for detailed performance descriptions). A relatively flat hexapod\cite{Symmetrie} provides the required sample fine motions (\textit{SampleX}, \textit{SampleY} and \textit{SampleZ}) along with roll (\textit{Chi}) and pitch (\textit{Eta}) while providing excellent stiffness. For angular rotations about the approximate surface normal, a \textit{Phi} goniometer riding on the hexapod motion is used to provide $\pm 180^{\circ}$ rotation of the sample inside the chamber. While a combination of the \textit{Phi} rotation and hexapod motions could rotate the sample about an arbitrary point on its surface, we were concerned about the practical stability of that during, for example, Bragg CDI imaging of a structure or defect on the sample. Thus, we added a larger, vertical-axis circle (\textit{Mu}) below the hexapod to provide precise rotation about the beam position for such measurements. In addition, the \textit{Mu} circle provides flexibility for incorporation of future instruments that, for example, might be too large to mount on the hexapod. Finally, to provide the required long travel in Z (e.g. to accommodate differences in incident beam height introduced by a focusing mirror), the \textit{Mu} goniometer is mounted on vertical motion stage (\textit{SystemZ}). 

The second diffractometer component is a detector mover with a relatively short working distance ($\sim$1 m) that has high stability and resolution. This is built on a large vertical-axis goniometer (\textit{Nu}) capable of handling a vertical load of 11000 N. The rotation \textit{Delta} of the detector about a horizontal axis is achieved through a compound motion provided by a long vertical translation stage (\textit{DelHeight}) mounted on \textit{Nu} at a radius $r_0 = 750$~mm from the sample, supporting a horizontal-axis goniometer (\textit{DelRot}). Virtual rotation to an angle $Delta = \Delta$ about the sample position is achieved by positioning the goniometer to $DelRot = \Delta$ and the translation to $DelHeight = r_0 \tan \Delta$ relative to the sample height. In order to change the distance between the sample and detector, an additional translation stage (\textit{Radial}) is mounted on the \textit{DelRot} goniometer, which is horizontal when \textit{DelRot} = 0. This allows adjustment of the angular resolution and total subtended angle of the area detector. It also allows optical components (e.g. those needed for full field imaging) to be moved close to the chamber for operation and retracted for sample changes and to avoid collisions with accessories during long motions of \textit{Nu} and \textit{Delta}. The \textit{Delta} tower is mounted on a manual rotation axis to allow for positioning the tower on either side of the sample centerline to provide flexibility in the addition of other characterization devices and, in particular, the simultaneous monitoring of sample reflectivity and Bragg diffraction. This motion is fixtured to allow easy and reproducible switching between the two configurations.

The errors in \textit{DelHeight} and \textit{DelRot}, $\epsilon_{DH}$ and $\epsilon_{DR}$, can contribute to errors in the combined \textit{Delta} motion; to first order, errors in \textit{Radial} do not contribute. To estimate this, we can express the \textit{Delta} value of the reference pixel on the area detector as
\begin{equation}
    \Delta = \tan^{-1} \left ( \frac{DelHeight + r \sin DelRot}{r_0 + r \cos DelRot} \right ),
\end{equation}
where $r$ is the radial position of the detector relative to the center of \textit{DelRot}. Differentiating gives the error in \textit{Delta} as
\begin{equation}
    \epsilon_{\Delta} = \frac{\epsilon_{DH} \cos^2 \Delta + \epsilon_{DR} \, r \cos \Delta}{r_0 + r \cos \Delta}.
\end{equation}
Using values of $\epsilon_{DH}$ and $\epsilon_{DR}$ from Tables I and II, and $\Delta = 0$ and $r$ = 250 mm, the accuracy and repeatability of \textit{Delta} are 0.0014$^{\circ}$ and 0.0007$^{\circ}$, respectively.

The sample mover and near detector mover are mounted together on a granite platform 1 m wide, 1.2 m long and 0.2 m thick. The granite platform has a vertical hole in its center with perpendicular holes through each side to allow for the introduction of utilities (e.g. pumping or electrical) on the vertical center of rotation of the sample mover. The granite block rests on six pads located on the APS's high stability concrete floor. The order of the rotations that connect the sample and detector reference frames to the laboratory reference frame is given in Table~\ref{table3}.

\begin{table}[htbp]
\caption{Order of rotations between laboratory and sample or detector reference frames.}
\begin{center}
\begin{ruledtabular}
\begin{tabular}{c c c}
Reference & Rotation & Axis Direction \\
Frame     &          & at Zero Angles \\
\hline
Sample & & \\
 & \textit{Phi}  & Z  \\
 & \textit{Chi}  & X  \\
 & \textit{Eta}  & Y  \\
 & \textit{Mu}   & Z  \\
Lab & & \\
\hline
Detector & & \\
 & \textit{Delta}  & Y  \\
 & \textit{Nu}   & Z  \\
Lab & & \\
\end{tabular}
\end{ruledtabular}
\end{center}
\label{table3}
\end{table}

The final envisaged component, a long-working-distance detector mover with precision appropriate to the pixel size of the detector (e.g. $\sim$50 microns) has not yet been built and is outside the scope of this paper. However, its future installation has been facilitated by not permanently mounting any critical components on the back wall of the hutch. In addition to these main components, a system to reduce incident backgrounds, provide x-ray attenuators and to monitor the incident x-ray intensity is mounted just upstream of the sample mover as shown in Fig.~\ref{figure1}.
 
\begin{figure}
    \includegraphics[width=\columnwidth]{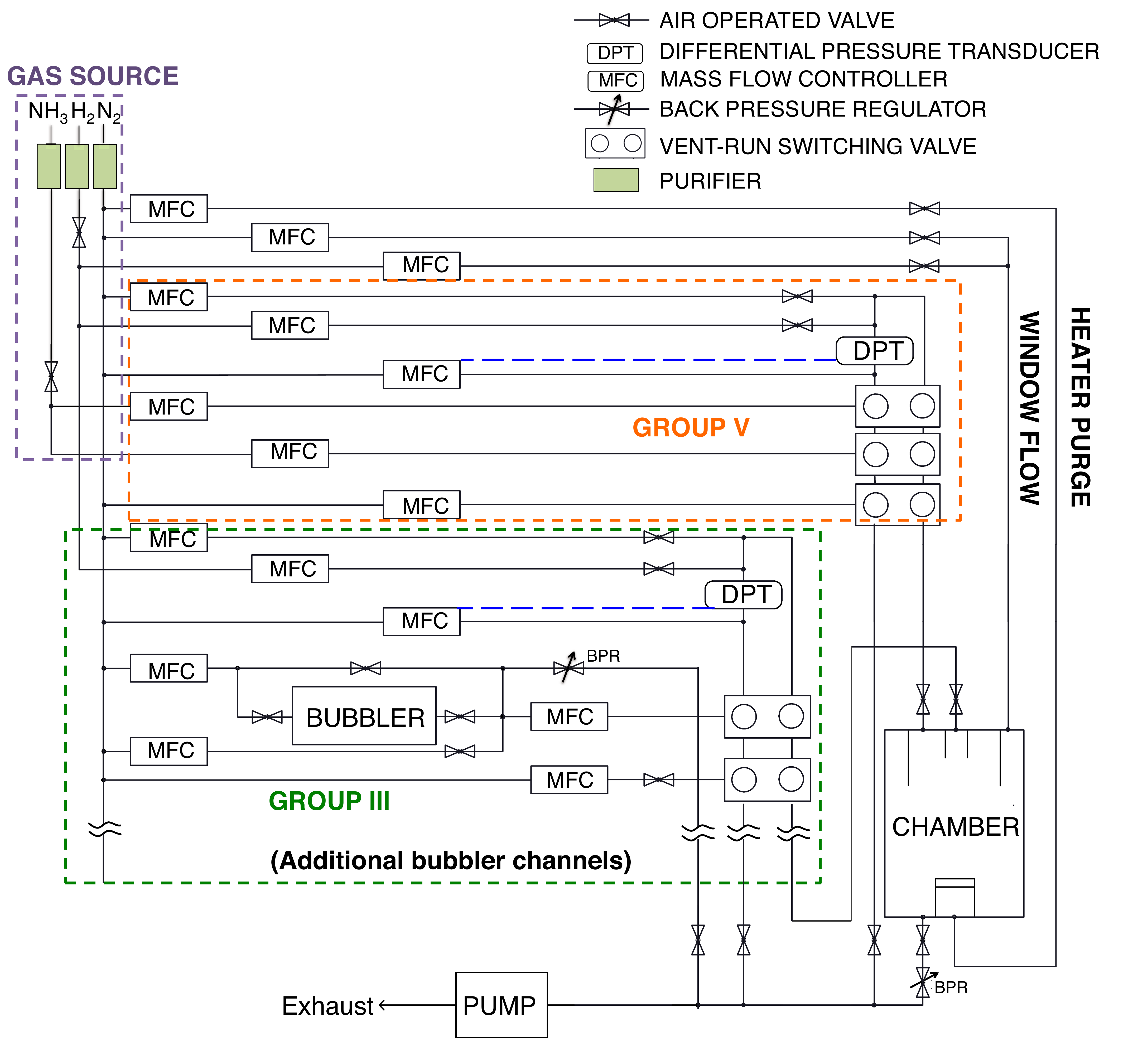} 
	\caption{\label{figure5} A schematic of the gas handling system, showing gas source, window and heater purge, group V, group III, and chamber sections. A typical example of one of the six group III channels is shown.}
\end{figure}

\subsection{Gas handling system}

Figure \ref{figure5} shows a schematic of the gas handling system. It consists of valves, mass flow controllers, and pressure controls for the gas sources, the purge, group V, and group III flows into the chamber, and the exhaust flows out of the chamber and vent lines. 

The group V precursor used is NH$_3$, while N$_2$ and H$_2$ mixtures can be used as carrier flows. Ventilated gas cabinets for the NH$_3$ and H$_2$ supply cylinders and purge manifolds, as well as a liquid N$_2$ dewar supplying N$_2$ carrier gas, are located just outside the shielded x-ray enclosure. Gas purifiers are used on each of these supply lines. Engineered safety features include automatic excess-flow shutoff valves for the NH$_3$ and H$_2$ supplies and redundant pressure regulators for the H$_2$ supply. The heater purge flow consists of N$_2$, while the window curtain flow can be a mixture of N$_2$ and H$_2$ to match the carrier gas for the group V and III flows.

Both the group V and group III sections include pressure-balanced process and vent lines, to allow switching of precursors to flow through or bypass the chamber without disturbing the stability of flow and pressure controllers. Equal make-up flows for each channel can be counter-switched simultaneously to keep the process and vent flows constant.

There are six group III channels available, with bubbler or compressed-gas precursor sources. This provides the flexibility to study the formation of alloys and heterointerfaces using different condensed-phase precursors such as triethyl gallium (TEGa), trimethyl indium (TMIn), trimethyl aluminum (TMAl), and dopants such as disilane gas. To provide a wide range of precursor flow rates into the chamber, on five channels the flow from the source can be diluted, with excess bypassing the chamber through a back pressure regulator. Most of the gas handling system, including the group III switching manifold and the group V sources, is located inside a large ventilated cabinet on the side wall of the shielded x-ray enclosure, several meters from the chamber. This allows full access to the chamber and room for motion of the detector goniometers. As shown in Fig.~\ref{figure1}, a small cabinet for the group III vent-run switching valves is mounted close to the deposition chamber. The short distance minimizes the delay between valve switching and change of group III vapor composition in the chamber, allowing rapid growth rate and film composition changes.

An automatic pressure regulator on the deposition chamber output flow is used to control the chamber pressure. All flows are exhausted into the chemical ventilation system for the beamline through a dry pump with chemically resistant seals. In addition to the ventilated enclosures housing most components of the gas handling system, a movable ventilation snorkel is located above the chamber to mitigate the risk of leaks at the chamber.

\subsection{Control System}

The valves, mass flow controllers, and pressure controls in the gas handling are interfaced to computers to control and monitor the vapor phase reactants. 
The interface consists of a programmable logic controller (PLC) with various sensors to monitor hazardous circumstances interlocked to automatically shut off gas flow if a hazard develops. Software control of the PLC, as well as the diffractometer motions, is implemented using EPICS\cite{EPICSwebsite} control system drivers to enable flexible and customizable measurement of growth processes. A dedicated graphical display is used to show the current state of valves, flows, pressures, and temperatures in the system, as well as the status of safety monitors and interlocks. A scriptable command-line interface implemented in the Python programming language is used to send individual commands or recipes to the system. The diffractometer control program, SPEC\cite{SPECwebsite} interfaces to the EPICS drivers and thus can control and record the gas handling system state during system operation.

All operations of the MOVPE system are carried out in conformance with a Standard Operating Procedures document. This not only mitigates safety risks but also improves the quality and reproducibility of results for experiments that often span multiple operators and personnel shifts.

\subsection{Safety Interlocks} 

The gas handling system includes a number of sensors interlocked to automatically place the system into a safe state if a fault condition is detected. Such a flow system abort generates an audible alarm, closes all mass flow controllers, and shuts all valves except the chamber exhaust, including the supply valves inside the H$_2$ and NH$_3$ gas cabinets. 

Gas concentration sensors are located in the gas handling system cabinet and near the chamber window for the detection of H$_2$ and NH$_3$ gas leaks. The H$_2$ sensors generate an abort when the concentration of H$_2$ exceeds 500 ppm, which is far below the flammability level. The NH$_3$ sensors generate an abort when the concentration exceeds the toxic level (25 ppm). Additional sensors monitor ventilation flow, pneumatic valve air supply pressure, power supply voltage, and PLC function, generating a gas system abort when a fault occurs. 

To mitigate the risk of a chamber overpressure in case its outlet valve is accidentally closed, the quartz window operating pressure (55 psig) is designed to exceed the gas supply pressures (40 psig). The chamber pressure is also limited by a burst disk (50 psig) and a mechanical check valve (1 psig) that vent into the exhaust system. Furthermore the chamber pressure monitor is interlocked to abort flows in the event of overpressure (3 psig).

When not in use, the flow system is locked into the abort state using a key switch. Manual abort buttons in the control room and near the chamber allow the operator to quickly abort flows in emergency situations. Additional sensors generate warnings, but not aborts, for certain other operational conditions (cabinet doors open, low water flow, low N$_2$ supply pressure).

In addition, both equipment and personnel protection are provided by hardware limit switches interlocked to the motor drivers of the diffractometer motions. In particular, to mitigate the pinch hazard between the \textit{Nu}-arm counterweight and the incident beam optics support, pressure plates covering the counterweight sides act as limit switches to stop motion.

\section{\label{sec:char}System characterization}

\subsection{\label{sec:charA}Computational fluid dynamics simulations}
\begin{figure} [ht!]
\includegraphics[width=\columnwidth]{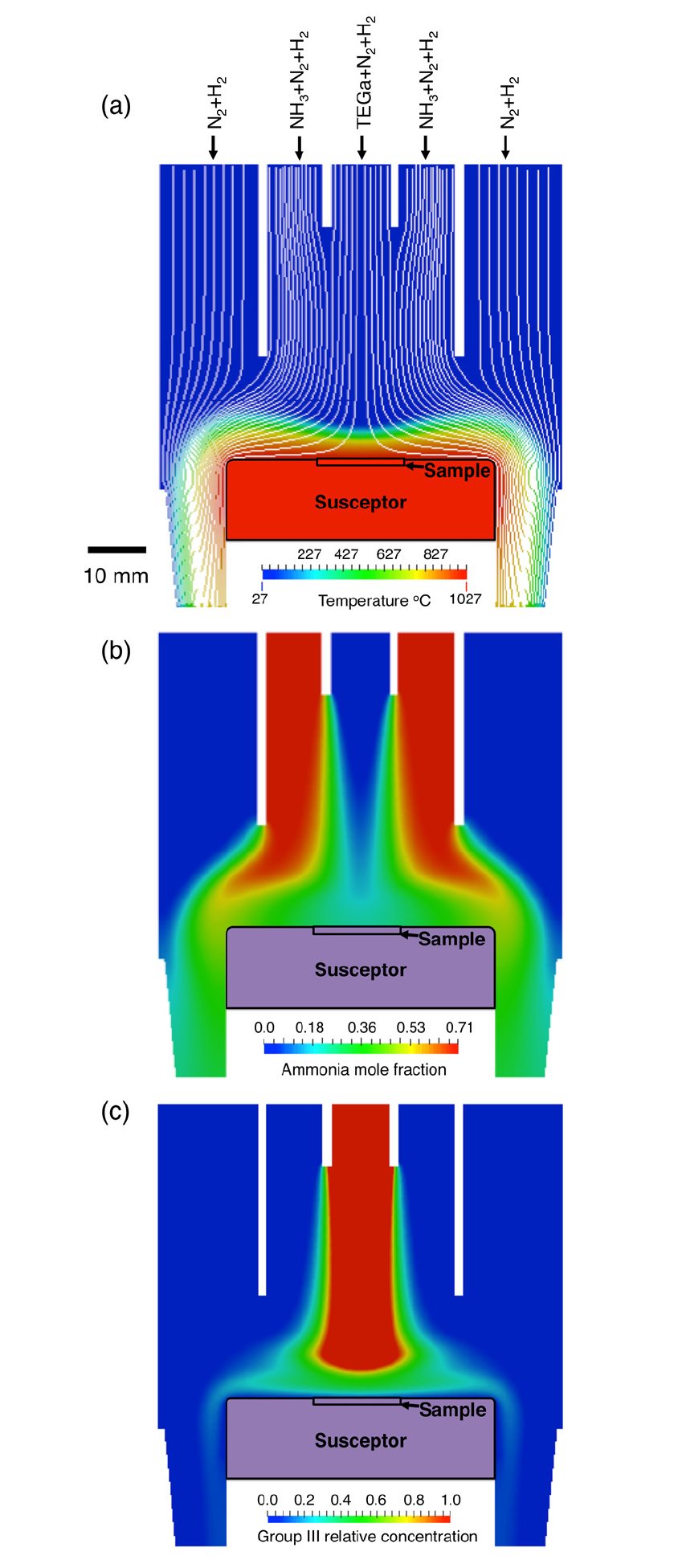} 
	\caption{\label{figure6} Results of a computational fluid dynamics analysis of the MOVPE chamber, for standard high-$T$ (1027 $^{\circ}$C, 50\% H$_2$ carrier) growth conditions. (a) Temperature distribution with flow streamlines; (b) NH$_3$ concentration distribution; (c) Group III concentration distribution. }
\end{figure} 

To optimize the chamber geometry and flow conditions, we performed computational fluid dynamics simulations as well as experimental tests. Figure \ref{figure6} shows the flow geometry in the chamber, with input flows entering from the top, and flowing down towards the sample and its mounting block, or ``susceptor''. It was found that the maximum total pressure without the formation of convection rolls at high $T$ was 200 Torr, for the maximum total flow rates compatible with overall system design. Optimized flows at 200 Torr are: 5.6 slpm in the outer window curtain flow, 3.8 slpm in the annular group V flow, and 0.9 slpm in the central group III flow. For typical high-NH$_3$ conditions, 2.7 slpm of the group V flow is NH$_3$, while the remainder of the carrier flows are either 100\% N$_2$ or a 50\% N$_2$ / 50\% H$_2$ mixture. Growth is controlled by the addition of very small fractions of group III precursors (e.g. TEGa) into the group III flow. 

Fig. \ref{figure6} shows typical simulated flow streamlines and temperature, NH$_3$, and group III distributions, obtained by self-consistently solving the mass, momentum, energy, and species (N$_2$, H$_2$, and NH$_3$) transport equations, as described in the literature. \cite{kee2005chemically} A time-dependent approach based on the PISO\cite{Ferzinger_cfd_1999} algorithm was used to solve the compressible Navier-Stokes equation, with the simulation proceeding until steady-state conditions were achieved. Since the flow of group III species are negligible compared with the group V and carrier gases, the transport of these species was modeled using the velocity and temperature profiles determined for the main constituents. The solver was implemented in OpenFOAM.\cite{weller1998tensorial}

The simulation domain comprises the whole inlet, window, sample, and downstream regions, although only the area close to the sample is shown here for clarity. This allowed us to model the development of the flow profiles within the three channels and determine the extent of upstream diffusion. A typical simulation consisted of 130,000 elements, though a finer mesh of 220,000 elements was used to ensure that the results were mesh-independent.

As shown in Fig.~\ref{figure6}(a), the flow follows a stagnation point pattern with a thermal boundary layer extending a distance of about 10 mm above the sample. The width of this layer is strongly dependent of the flow in the inner group III channel, decreasing with increasing flow. This flow dependence explains the extension of the thermal boundary layer near the edge of the susceptor, as the flow velocity in the group V channel is typically smaller than that in the group III. 

The group III flow also has a strong impact on the NH$_3$ distribution, Fig.~\ref{figure6}(b), which has a minimum at the center of the chamber. In contrast, the concentration of group III species has a maximum at the center of the chamber, as shown in Fig.~\ref{figure6}(c). The results shown here are obtained assuming that all group III precursor arriving at the susceptor is deposited, an assumption consistent with the growth being transport-limited for the 
group III precursor.

\subsection{Thermal expansion of the sample mount}

A primary design feature of the chamber is the use of a fused quartz tube mount to minimize sample motion when changing temperature. We observed that the change in the height of the sample when heating it from room temperature to 1000 $^\circ$C was about 50 $\mu$m. This agrees with the calculated thermal expansion contributions from the supporting fused quartz tube ($\sim40 \mu$m) and the molybdenum susceptor ($\sim10 \mu$m), assuming a linear temperature distribution in the tube. (It is possible that the lower section of the tube remains cooler, but that expansion of the stainless steel base contributes instead.) Once the temperature distribution has stabilized, e.g. after about 30 minutes, the sample position is stable to the sub-micron precision needed for coherent x-ray measurements.

\subsection{\label{sec:temperaturecalibration}Temperature calibration}

To avoid contamination of the MOVPE environment, our samples typically simply rest in a shallow indentation on the top surface of the molybdenum susceptor with no thermal conduction compound or other material to assist in equalizing the sample and susceptor temperatures. While a thermocouple embedded in the susceptor monitors its temperature, the sample temperature can be significantly lower because of contact with the gas environment, as well as radiative cooling. This differential temperature is affected by chamber conditions that alter conductive, convective, and radiative heat transport, such as total chamber pressure, carrier gas flow and composition, and deposit buildup on susceptor. In order to calibrate the sample temperature, we periodically measure this differential under various typical chamber conditions using an optical interference technique\cite{1991_Saenger_ApplOpt30_1221} and a reference sample.

\begin{figure}[ht!]
	\includegraphics[width=\columnwidth]{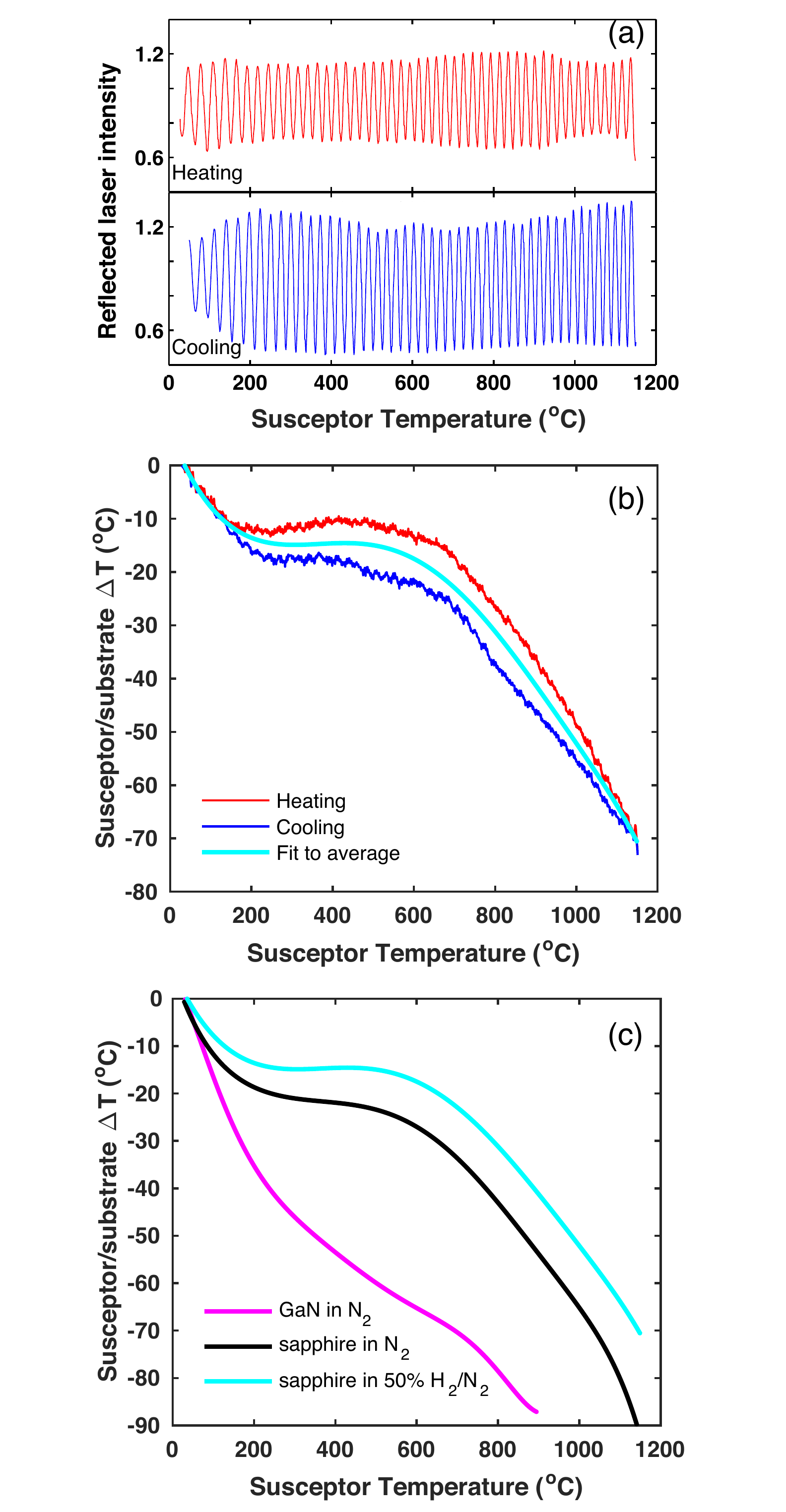}
	\caption{\label{figure7} Plots (a) and (b) illustrate the temperature calibration procedure using a sapphire $(0001)$ reference sample under 50\% H$_2$ flows typical for high-temperature GaN growth. (a) Interference oscillations measured during heating and cooling cycles. (b) Extracted temperature differential between sample and susceptor as a function of susceptor $T$ during heating and cooling, and a polynomial fit to the average. (c) Comparison of $T$ calibrations using sapphire with and without H$_2$ in the carrier gas, and a calibration using GaN (0001) in pure N$_2$ carrier.}
\end{figure}

Using a reference sample for which the index of refraction and thickness versus temperature are known, measurement of the interference fringes in the normal-incidence laser reflection from the upper and lower surfaces is a robust method to determine the sample temperature.\cite{1991_Saenger_ApplOpt30_1221} For this purpose, an external diode laser beam (wavelength $\lambda = 633$~nm) enters the chamber via a lens at the on-axis port (21) in Fig. \ref{figure3} that focuses at the sample position. The reflected signal is measured with a photodiode. As the sample expands or contracts during heating or cooling, the measured intensity oscillates. Fig.~\ref{figure7}(a) shows typical intensity oscillations measured during a cycle of heating and cooling at 0.5 $^{\circ}$/s for a $(0001)$ sapphire sample with both surfaces polished. Flow conditions were typical for high-$T$ growth (given in Section~\ref{sec:charA}), with 50\% N$_2$ / 50$\%$ H$_2$ carrier gas. 

The phase change $\Delta \phi$ of the oscillations can be related to the average sample temperature $T$ by 
\begin{equation}
\frac{\Delta\phi}{2 \pi} = 2 n(T)\frac{c(T)}{c_{0}}\frac{L_{0}}{\lambda},
\label{eq:Dphi}
\end{equation}
where $n(T)$ is the refractive index as a function of $T$, $c(T)/c_0$ is the fractional change in the lattice parameter as a function of $T$, and $L_{0}$ is the thickness of the reference sample at ambient $T$. Accurate values of $n(T)$ and $c(T)$ for sapphire are available in the literature.\cite{1986_Tapping_JOSAA3_610,1977_Touloulian_TPRCseries13} For our reference sample with $L_0 = 0.439$~mm, the intensity goes through a full oscillation for a temperature change of about 20 $^{\circ}$C. Since it is straightforward to extract the phase to $\sim 1/20$ of an oscillation, the precision of the method is $\sim$ 1 $^{\circ}$C; however, its accuracy depends primarily on how well $n(T)$ and $c(T)$ are known.

With the assumption that the susceptor and sample temperatures are identical at ambient prior to heating, the sample temperature can be extracted from the measured phase change by inverting Eq.~(\ref{eq:Dphi}). Fig.~\ref{figure7}(b) shows the results of a typical calibration of the sapphire sample $T$ as a function of susceptor $T$, plotted as the temperature differential, for both heating and cooling. The temperature differential is smaller in magnitude during heating than during cooling. We use a polynomial fit to the average of the heating and cooling values (smooth central curve in Fig.~\ref{figure7}(b)) to give the calibration for the sample $T$ after equilibration at constant susceptor $T$. 

The above analysis neglects temperature gradients through the thickness of the reference sample. This gradient can be estimated by a thermal analysis. Of the typical total heater power of 550 W needed to reach a susceptor $T$ of 1100 $^{\circ}$C (sample $T$ of 1025 $^{\circ}$C), one can estimate that 150 W is radiated and 400 W is conducted into the gas surrounding the susceptor. This gives a thermal flux conducted through the sample of 0.08 W/mm$^2$, assuming uniform flux from the top, bottom, and side surfaces of the susceptor. The thermal conductivity of sapphire \cite{2014_Hofmeister_PhysChemMin41_361} at 1025 $^{\circ}$C is 7.5 Wm$^{-1}$K$^{-1}$, giving a temperature difference of about 5 $^{\circ}$C through the thickness of the sapphire reference sample.

Figure \ref{figure7}(c) compares calibrations obtained from the sapphire sample for carrier gas compositions of 50\% N$_2$ / 50\% H$_2$ and pure N$_2$. The presence of hydrogen typically reduces the magnitude of the differential by about 10 K. 

We also carried out a calibration using a {0.354-mm-thick} GaN $(0001)$ reference sample, using literature values for the refractive index \cite{watanabe2008temperature,2006_Liu_physstatsolc3_1884} and lattice parameter\cite{2000_Reeber_JMR15_40} as functions of $T$. The calculated GaN sample $T$ is significantly lower that of sapphire, at the same susceptor $T$. Since the thermal conductivity of single-crystal GaN \cite{2007_Shibata_MaterialsTrans48_2782} is more than an order of magnitude larger than that of sapphire, this would not explain its lower temperature. The cooling effect may come from the endothermic decomposition of NH$_3$, which is expected to be more efficient on GaN than on sapphire. Based on the enthalpy of reaction,\cite{1998_Chase_JPCRDMono9} decomposition of the full 2.7 slpm NH$_3$ flow will absorb 112 W of power, which could significantly affect the temperature distribution. Further experiments will be required to understand the difference between the sapphire and GaN temperature calibrations. Since the temperature-dependent refractive index of sapphire is better established than that of GaN, we use the sapphire calibrations for the temperatures reported here.

\subsection{Growth uniformity and efficiency}

\textit{In situ} optical interferometry can also be used to monitor film growth rates,\cite{1999_Stephenson_MRSBull24_21} a capability that is very useful to optimize MOVPE conditions in preparation for synchrotron x-ray experiments. During GaN growth, interference oscillations occur with the addition of every 133 nm (${\lambda}/{2n}$) to the film thickness. It is convenient to distinguish oscillations due to growth from those due to temperature changes by using a thin film on a substrate of different refractive index. We typically use a thin film of GaN on a sapphire substrate for laser monitoring of growth. Interference between reflections from the upper and lower surfaces of the film produce oscillations when film thickness changes due to growth. Using a substrate with an unpolished back surface avoids superimposing any oscillations due to small temperature variations of the substrate. 

\begin{figure} [ht]
	\includegraphics[width=\columnwidth]{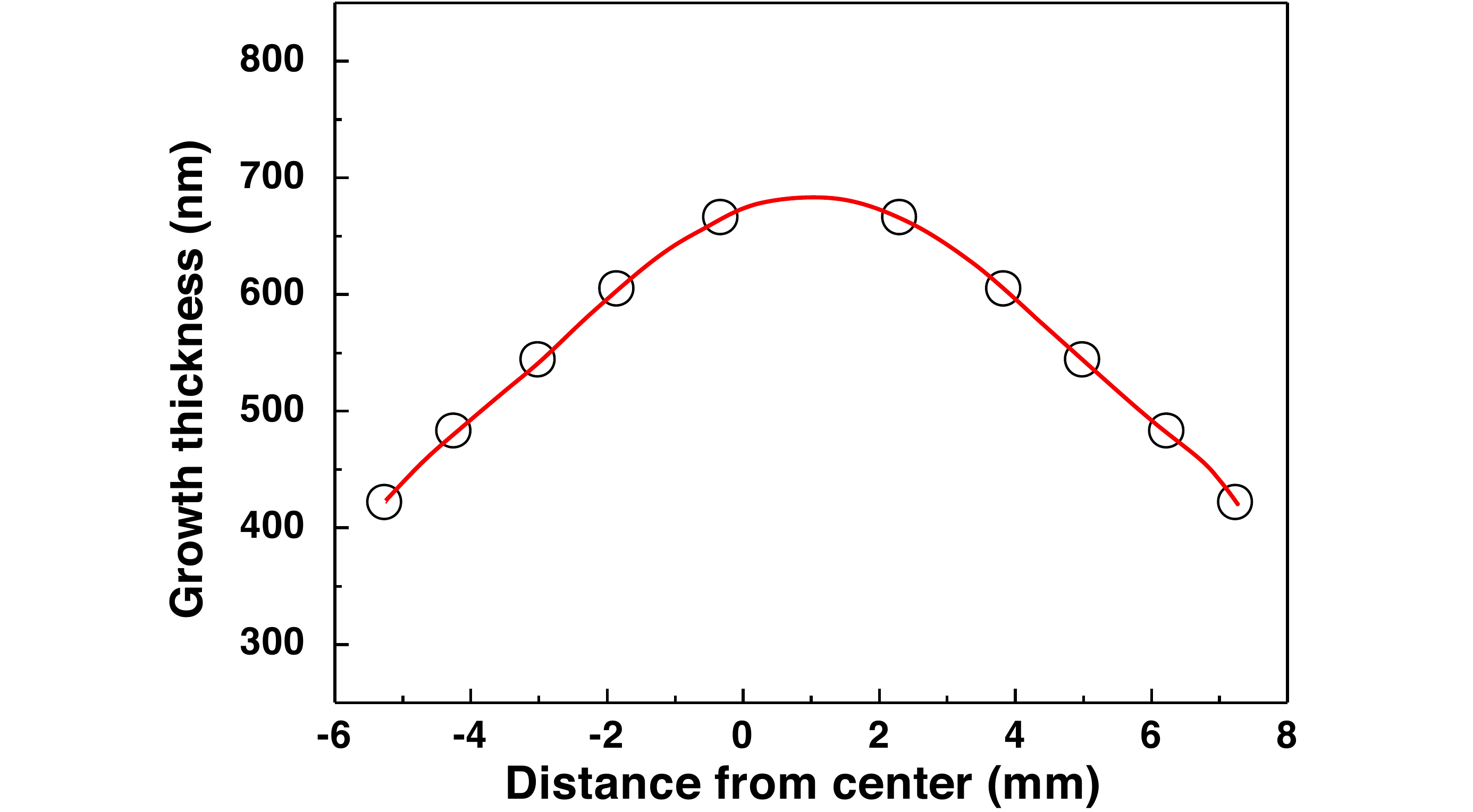}
	\caption{\label{figure8} Growth thickness non-uniformity developed after depositing approximately 700 nm of GaN at 0.25 nm/s under standard conditions at 1036 $^{\circ}$C. Points show positions of minima and maxima of fringe pattern; line is guide to the eye.}
\end{figure}

After growth of a film sufficiently thick to produce several optical fringes at the center of the sample, the thickness uniformity can be determined by characterizing the pattern of fringes across the sample surface. Our chamber geometry with optimized flows typically produces a maximum in the growth rate near the center of the sample. Fig.~\ref{figure8} shows the typical deposition nonuniformity observed. Asymmetry in the flow pattern due to non-concentricity of the sample mount and the chamber wall leads to the growth maximum being typically 1 mm off center. The absolute growth amount in the center is determined using the \textit{in situ} optical monitor, while the distribution relative to this location is determined from the fringe pattern. To observe the fringe pattern it is convenient to use a fluorescent light with dominant wavelengths of 545 and 611 nm, giving $\sim 122$ nm of film thickness per fringe. The x-ray beam typically illuminates a region on the sample smaller than 2 mm, so the growth rate is uniform to within $\sim 10$\% for typical \textit{in situ} x-ray experiments. 

Under a wide range of GaN MOVPE conditions, the growth rate is proportional to the input flow rate of the Ga precursor, independent of temperature. A TEGa supply of 1 $\mu$mol/min gives a growth rate near 0.1 nm/s. We have measured this linearity over growth rates from 0.001 to 1 nm/s. One can determine the overall efficiency of growth by dividing the number of moles of GaN deposited, based on the distribution in Fig.~\ref{figure8}, by the number of moles of precursor supplied. We observe an overall efficiency of $\sim 9$\%. 

\section{X-ray scattering performance}

The performance of the chamber for \textit{in situ} surface x-ray scattering under growth conditions, including real-time monitoring during crystal growth, has been systematically studied. We have also demonstrated the precision and reproducibility of sample positioning using microdiffraction from a patterned substrate, and chamber stability sufficient for coherent x-ray scattering. Measurements were made using x-ray energies in the range 24 to 26 keV.

\subsection{Wide-angle surface x-ray scattering}

\begin{figure}
	\includegraphics[width=\columnwidth]{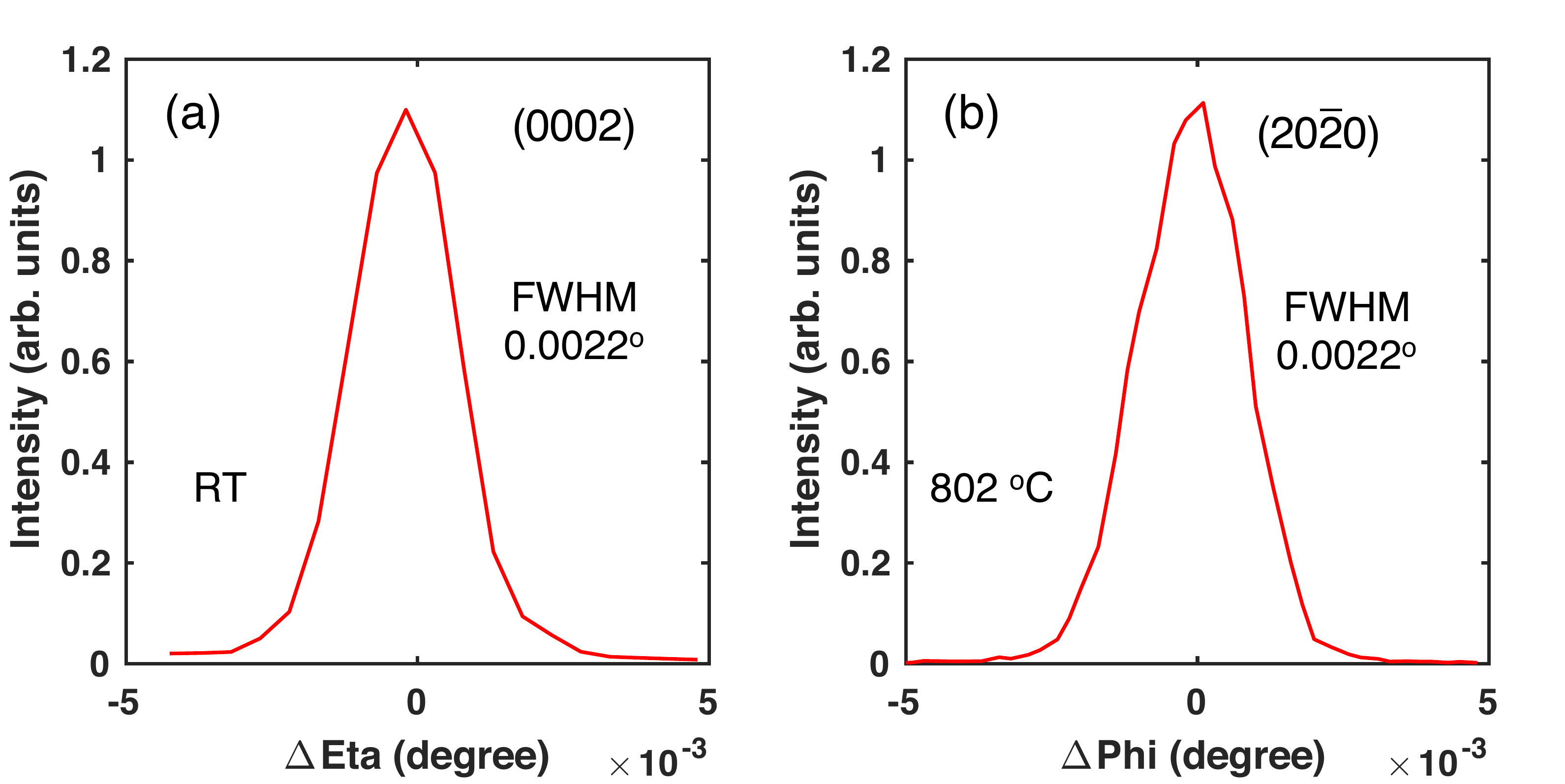}
	\caption{\label{figure9} Rocking curves of GaN single crystals with $(0001)$ surface orientations. (a) Out-of-plane $(0002)$ reflection at room temperature. (b) In-plane $(20\overline{2}0)$ reflection at 802 $^{\circ}$C.}
\end{figure}

To demonstrate precision of sample rotation, Fig. \ref{figure9} shows measurements of the rocking curves of out-of-plane and in-plane Bragg peaks of GaN crystals with $(0001)$ surface orientations. The 0.0022$^{\circ}$ full-width at half maximum (FWHM) of the $(0002)$ reflection demonstrates the \textit{Eta} motion of the hexapod, while the 0.0022$^{\circ}$ FWHM of the $(20\overline{2}0)$ reflection demonstrates the \textit{Phi} motion and the rotary feedthrough of the chamber.

\begin{figure}
	\includegraphics[width=\columnwidth]{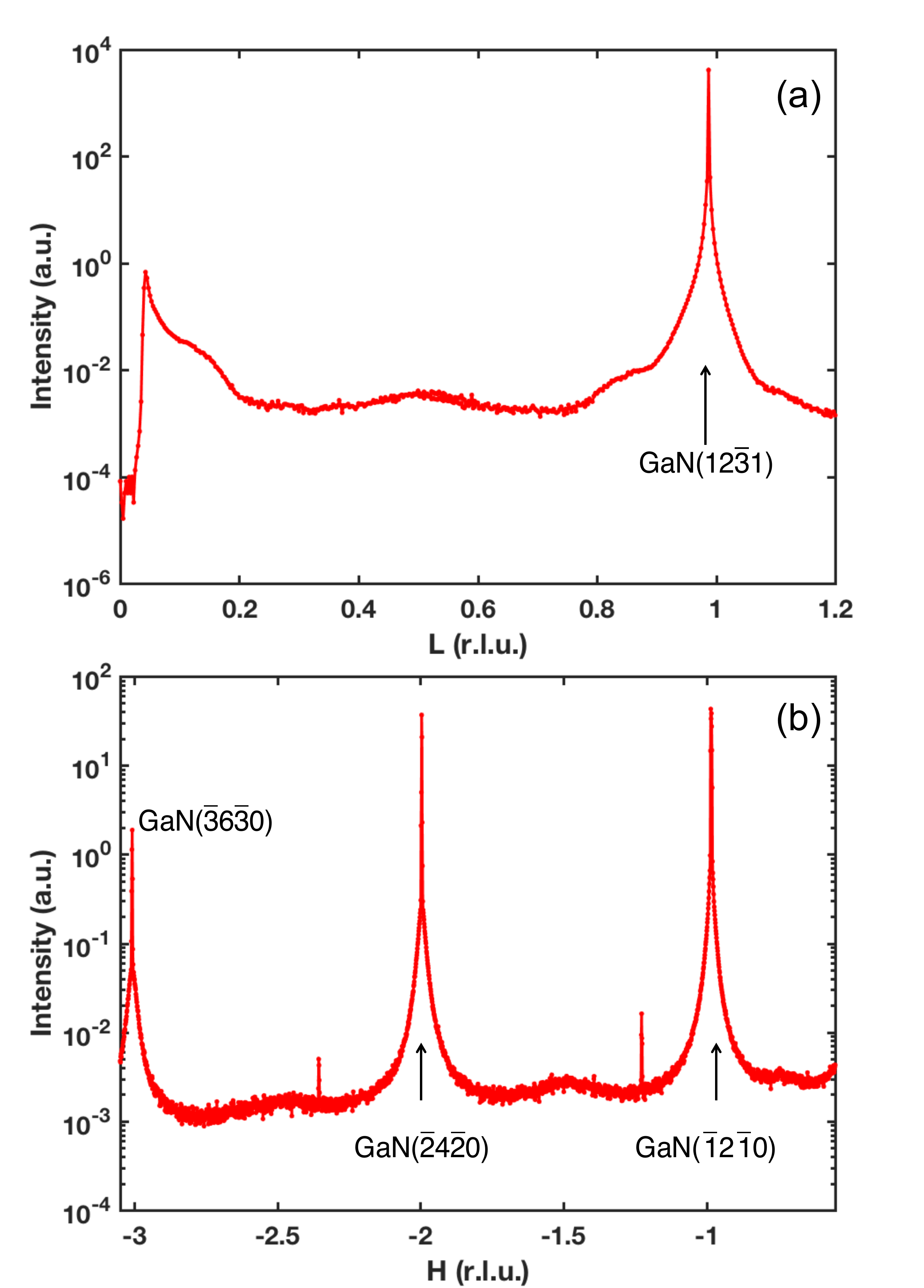}
	\caption{\label{figure10} X-ray scattering from a GaN $(20\overline{2}1)$ semipolar surface at 476 $^\circ$C under standard NH$_3$ flow with 100\% N$_2$ carrier. (a) Crystal truncation rod through $(\overline{1}2\overline{1}0)$, scanning along the $(\text{2L}, 0, \overline{\text{2L}}, \text{L})$ surface normal. Incident angle fixed at $\alpha=1^{\circ}$. (b) In-plane scattering along the $(\overline{1}2\overline{1}0)$ azimuth with $\alpha=0.14^\circ$. Reciprocal lattice units are based on room-temperature GaN lattice parameters.}	
\end{figure}

Figure \ref{figure10} shows examples of surface-sensitive scattering from a GaN single crystal with a semipolar $(20\overline{2}1)$ surface orientation. The sample was at 476 $^\circ$C under standard N$_2$ and NH$_3$ flows. Fig.~\ref{figure10}(a) shows a CTR, which is a streak of scattering extending normal to the crystal surface.\cite{1990_Fuoss_AnnRevMatSci20_365} Fig.~\ref{figure10}(b) shows an in-plane scan measured at grazing incidence. Both scans demonstrate the ability to follow a desired path in reciprocal space by motion of multiple angles, with the incident angle $\alpha$ kept fixed. 

\subsection{Real-time monitoring of growth}

\begin{figure}
    \centering
    \includegraphics[width=\columnwidth]{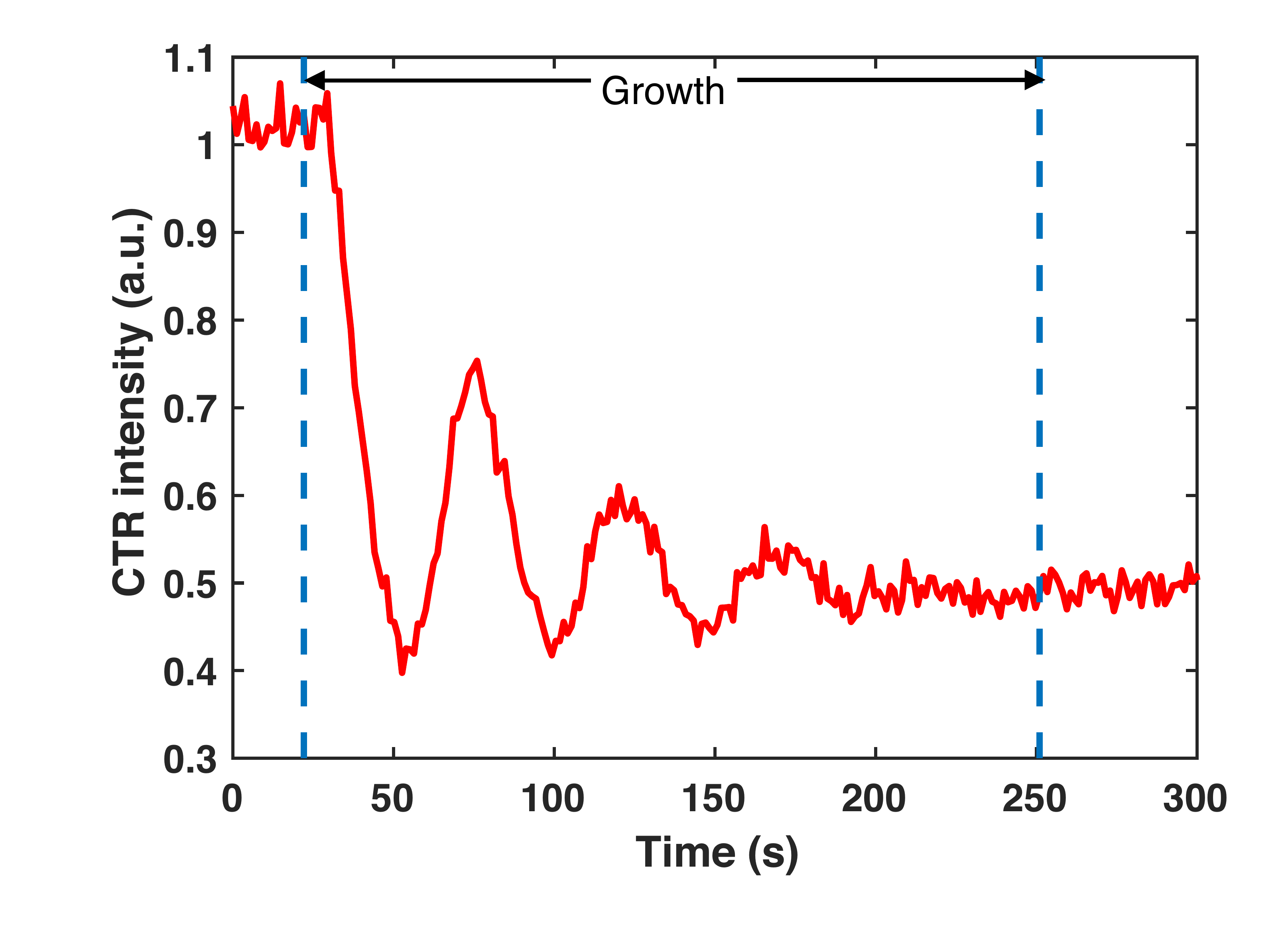}
    \caption{\label{figure11} Homoepitaxial growth oscillations of a GaN single crystal with $(10\overline{1}0)$ surface orientation, measured on the ($\overline{1}$+$\Delta \text{H}$, 2, $\overline{1}$-$\Delta \text{H}$, 0) CTR at $\Delta \text{H}$ = 0.5 using a 3 $\mu$m$\times$3 $\mu$m coherent x-ray beam. Input flow rate of TEGa was 0.05 $\mu$mol/min, $T$ = 666 $^{\circ}$C. Vertical lines denote inject and vent of TEGa.} 
\end{figure}

Using an earlier generation growth chamber and system, we previously mapped the transitions between homoepitaxial growth modes on GaN single crystals with $(10\overline{1}0)$ surface orientations.\cite{2014_Perret_APL105_051602} Figure~\ref{figure11} shows a real-time measurement of oscillations in the CTR intensity during homoepitaxial growth under similar conditions using the new chamber. Each oscillation corresponds to the growth of one monolayer of the crystal. In this case, measurements were made with a transversely coherent beam. This verifies that we can carry out dynamic growth mode studies with coherent x-ray illumination using the new system.


\subsection{Reproducibility of positioning - Microdiffraction}

\begin{figure}
	\includegraphics[width=\columnwidth]{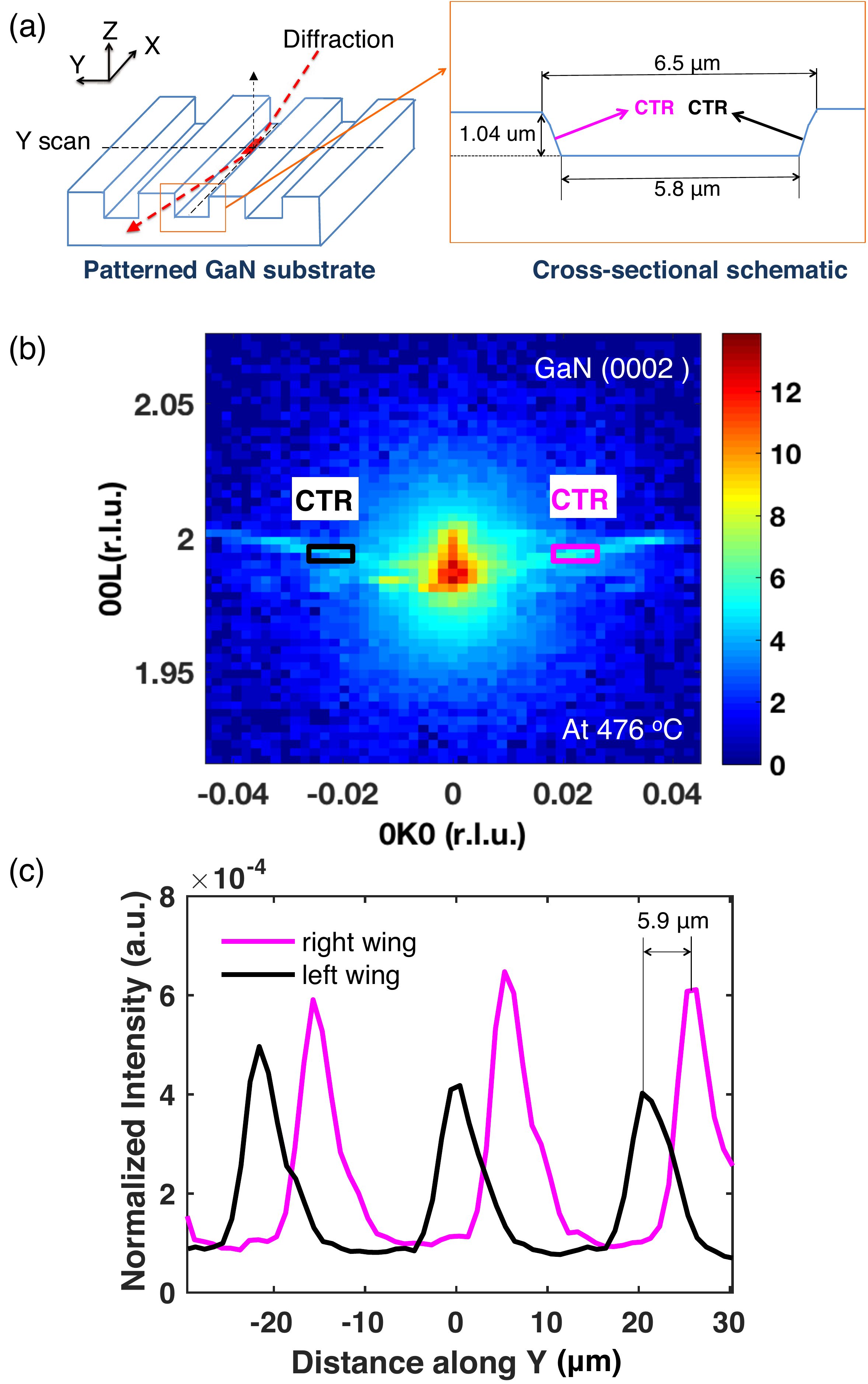}
	\caption{\label{figure12} Microdiffraction using an incident beam size of 5 $\mu$m(Y)$\times$50 $\mu$m(X). (a) The sample was a GaN single crystal with (0001) surface at 476 $^\circ$C, patterned with an array of micron-scale trenches. The cross section shows the dimensions of each trench. (b) Diffraction distribution near GaN $(0002)$ with incident beam in the center of a trench. Two CTRs with angles $\pm18^{\circ}$ from horizontal arise from the two side walls of the trench. (c) Intensity of each CTR as the sample position is scanned normal to the trench direction. The different variations with position indicate the spatial resolution of the measurement.}
\end{figure}

To demonstrate that the chamber can be used to study surface morphology variations on a micron length scale, microdiffraction measurements were made from a patterned GaN sample, as shown in Fig.~\ref{figure12}(a). The incident beam size was 5 $\mu$m(Y)$\times$50 $\mu$m(X), and the sample was at $T$ = 476 $^\circ$C under typical NH$_3$ and N$_2$ flows. Fig.~\ref{figure12}(b) shows the diffracted intensity distribution near the GaN $(0002)$ Bragg peak measured on an area detector, with the incident beam centered on a surface trench. Two CTRs can be seen with angles $\pm18^{\circ}$ from horizontal, arising from the two tilted side walls of the trench. By monitoring each CTR as the sample position is scanned, as shown in Fig.~\ref{figure12}(c), we can separate the scattering from each side wall and characterize the spatial resolution of the measurement. The 5.9 nm separation of the peaks in Fig.~\ref{figure12}(c) is consistent with trench width measured post-growth with scanning electron microscopy. The results indicate that the system has the high position resolution and angular stability needed for microdiffraction studies.

\subsection{Stability of system for coherent x-ray scattering}

To more stringently test the stability of the chamber and diffractometer for experiments with coherent x-ray beams, we measured speckle patterns from surface steps on a miscut GaN (0001) sample. The sample was at room temperature, and had a surface orientation 0.29$^{\circ}$ away from (0001). Measurements were made near the (000L) CTR at $\text{L} = 0.9$. A 24.05 keV ($\lambda = 0.0516$ nm) partially coherent beam of diameter $w$ = 5 $\mu$m was used, giving an expected speckle size of $s = 0.886 \lambda/w$ = 9 $\mu$rad.\cite{sutton1991observation} Using a CCD area detector with 20 $\mu$m pixels located 2.33 m from sample on the back wall of the hutch, each pixel subtended approximately one speckle.

\begin{figure}
	\includegraphics[width=\columnwidth]{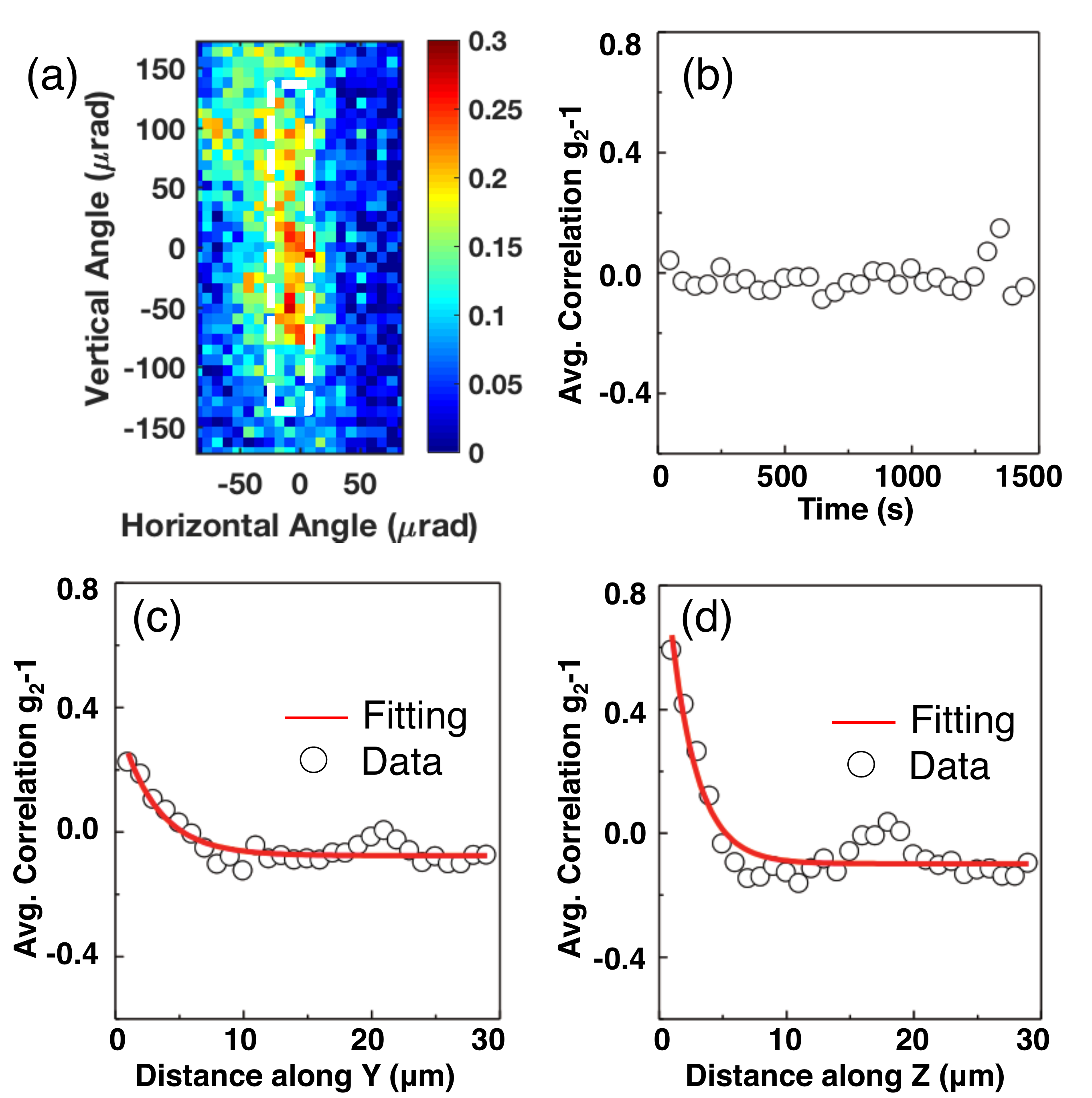}
	\caption{\label{figure13} Initial coherent x-ray characterization of a GaN (0001) sample at room temperature, using a 24.05 keV incident beam of 5 $\mu$m diameter. (a) Time-averaged speckle at fixed sample position, showing region of pixels used for autocorrelation. (b) Autocorrelation from a time series at fixed position, consistent with a static speckle pattern. (c) and (d) Autocorrelation from scans of the sample in the Y and Z directions. Fits give x-ray coherence lengths of 2.8 $\mu$m and 2.1 $\mu$m in the horizontal (Y) and vertical (Z), respectively.} 
\end{figure}

After normalizing to the incident intensity, we analyzed the normalized intensity fluctuations $I(t)$ in each pixel using the standard XPCS autocorrelation function,\cite{2014_Shpyrko_JSynchRad21_1057}   
\begin{equation}\label{eq:correlation}
g_2(\Delta t) - 1 = \frac{\langle I(t) I(t+\Delta t)\rangle_t - \langle I\rangle_t^2}{\langle I\rangle_t^2}.
\end{equation}
Fig.~\ref{figure13}(a) shows the autocorrelation from a time series measured at a fixed location on the sample. The flat signal indicates that the speckle pattern is static on the 1500~s time scale, indicating that the chamber and diffractometer are very stable. Fig. \ref{figure13}(b) and (c) show the autocorrelation function obtained when the sample is scanned along the Y and Z directions, respectively. In this case, the time variable in Eq.~\ref{eq:correlation} corresponds to a spatial variable, and the speckle pattern is expected to evolve with a correlation length equal to the transverse coherence length of the x-ray beam. Fits to an exponential function (red curves) give coherence lengths of 2.8 $\mu$m and 2.1 $\mu$m in horizontal (Y) and vertical (Z) directions, respectively, which are in reasonable agreement with calculated values based on the beamline source size and focusing optics. The small peak away from zero is likely an artifact due to insufficient averaging because of the limited range of the scan. These preliminary results indicate that the system is sufficiently stable to observe the time dependence in speckle patterns from GaN surfaces under MOVPE conditions, enabling XPCS measurements of atomic-scale dynamics.

\section{Summary and outlook}

Despite major advances in the development of devices using III-nitride materials, many fundamental questions remain regarding the synthesis of these metastable materials that could bring applications to a new level. The unique features of this instrument will bring to bear advanced coherent x-ray techniques (surface XPCS, CDI, and microbeam diffraction) to reveal the atomic-scale processes occuring during MOVPE growth of these materials. 

Initial studies planned include determining step dynamics in the MOVPE environment, both at equilibrium and during step-flow growth. The dependence of step, kink, and adatom energies and dynamics on surface and step orientation has a fundamental impact on crystal growth mechanisms. Yet standard incoherent x-ray methods have been blind to these processes, because the average structure remains constant at equilibrium or during steady-state step-flow growth. Surface XPCS\cite{pierce2009surface} will reveal these dynamics. In addition, XPCS studies of two-time correlations\cite{malik1998coherent} will provide a qualitatively new view of non-steady-state processes, such as layer-by-layer growth, revealing the effects of surface dynamics on correlations in island nucleation sites on subsequent atomic layers. These studies will greatly enhance our fundamental understanding MOVPE and allow for the development of improved synthesis methods for nitride materials, helping to enable the fabrication of next-generation devices with atomic-level control. 

\begin{acknowledgments}
We would like to thank Mark Rivers, Matthew Newville, and Dohn Arms for their expert advice and help in creating the EPICS interfaces for the gas handling system and hexapod.
This research was supported by the U.S. Department of Energy (DOE), Office of Science, Office of Basic Energy Sciences, Division of Materials Science and Engineering. This research used resources of the Advanced Photon Source, a DOE Office of Science User Facility operated for the DOE Office of Science by Argonne National Laboratory under Contract No. DE-AC02-06CH11357.
\end{acknowledgments}

\input{GJu_MOVPE_apparatus_v6revised1.bbl}

\end{document}

%% file: GJu_MOVPE_apparatus_v6revised1.bbl
%

%% file: GJu_MOVPE_apparatus_v6revised1.bbl
\begin{thebibliography}{43}%
\makeatletter
\providecommand \@ifxundefined [1]{%
 \@ifx{#1\undefined}
}%
\providecommand \@ifnum [1]{%
 \ifnum #1\expandafter \@firstoftwo
 \else \expandafter \@secondoftwo
 \fi
}%
\providecommand \@ifx [1]{%
 \ifx #1\expandafter \@firstoftwo
 \else \expandafter \@secondoftwo
 \fi
}%
\providecommand \natexlab [1]{#1}%
\providecommand \enquote  [1]{``#1''}%
\providecommand \bibnamefont  [1]{#1}%
\providecommand \bibfnamefont [1]{#1}%
\providecommand \citenamefont [1]{#1}%
\providecommand \href@noop [0]{\@secondoftwo}%
\providecommand \href [0]{\begingroup \@sanitize@url \@href}%
\providecommand \@href[1]{\@@startlink{#1}\@@href}%
\providecommand \@@href[1]{\endgroup#1\@@endlink}%
\providecommand \@sanitize@url [0]{\catcode `\\12\catcode `\$12\catcode
  `\&12\catcode `\#12\catcode `\^12\catcode `\_12\catcode `\%12\relax}%
\providecommand \@@startlink[1]{}%
\providecommand \@@endlink[0]{}%
\providecommand \url  [0]{\begingroup\@sanitize@url \@url }%
\providecommand \@url [1]{\endgroup\@href {#1}{\urlprefix }}%
\providecommand \urlprefix  [0]{URL }%
\providecommand \Eprint [0]{\href }%
\providecommand \doibase [0]{http://dx.doi.org/}%
\providecommand \selectlanguage [0]{\@gobble}%
\providecommand \bibinfo  [0]{\@secondoftwo}%
\providecommand \bibfield  [0]{\@secondoftwo}%
\providecommand \translation [1]{[#1]}%
\providecommand \BibitemOpen [0]{}%
\providecommand \bibitemStop [0]{}%
\providecommand \bibitemNoStop [0]{.\EOS\space}%
\providecommand \EOS [0]{\spacefactor3000\relax}%
\providecommand \BibitemShut  [1]{\csname bibitem#1\endcsname}%
\let\auto@bib@innerbib\@empty
\bibitem [{\citenamefont {Stephenson}\ \emph {et~al.}(1999)\citenamefont
  {Stephenson}, \citenamefont {Eastman}, \citenamefont {Auciello},
  \citenamefont {Munkholm}, \citenamefont {Thompson}, \citenamefont {Fuoss},
  \citenamefont {Fini}, \citenamefont {DenBaars},\ and\ \citenamefont
  {Speck}}]{1999_Stephenson_MRSBull24_21}%
  \BibitemOpen
  \bibfield  {author} {\bibinfo {author} {\bibfnamefont {G.~B.}\ \bibnamefont
  {Stephenson}}, \bibinfo {author} {\bibfnamefont {J.~A.}\ \bibnamefont
  {Eastman}}, \bibinfo {author} {\bibfnamefont {O.}~\bibnamefont {Auciello}},
  \bibinfo {author} {\bibfnamefont {A.}~\bibnamefont {Munkholm}}, \bibinfo
  {author} {\bibfnamefont {C.}~\bibnamefont {Thompson}}, \bibinfo {author}
  {\bibfnamefont {P.~H.}\ \bibnamefont {Fuoss}}, \bibinfo {author}
  {\bibfnamefont {P.}~\bibnamefont {Fini}}, \bibinfo {author} {\bibfnamefont
  {S.~P.}\ \bibnamefont {DenBaars}}, \ and\ \bibinfo {author} {\bibfnamefont
  {J.~S.}\ \bibnamefont {Speck}},\ }\href {\doibase 10.1557/S088376940005168X}
  {\bibfield  {journal} {\bibinfo  {journal} {MRS Bulletin}\ }\textbf {\bibinfo
  {volume} {24}},\ \bibinfo {pages} {21} (\bibinfo {year} {1999})}\BibitemShut
  {NoStop}%
\bibitem [{\citenamefont {Jiang}\ \emph {et~al.}(2006)\citenamefont {Jiang},
  \citenamefont {Wang}, \citenamefont {Munkholm}, \citenamefont {Streiffer},
  \citenamefont {Stephenson}, \citenamefont {Fuoss}, \citenamefont {Latifi},\
  and\ \citenamefont {Thompson}}]{2006_Jiang_APL89_161915}%
  \BibitemOpen
  \bibfield  {author} {\bibinfo {author} {\bibfnamefont {F.}~\bibnamefont
  {Jiang}}, \bibinfo {author} {\bibfnamefont {R.-V.}\ \bibnamefont {Wang}},
  \bibinfo {author} {\bibfnamefont {A.}~\bibnamefont {Munkholm}}, \bibinfo
  {author} {\bibfnamefont {S.~K.}\ \bibnamefont {Streiffer}}, \bibinfo {author}
  {\bibfnamefont {G.~B.}\ \bibnamefont {Stephenson}}, \bibinfo {author}
  {\bibfnamefont {P.}~\bibnamefont {Fuoss}}, \bibinfo {author} {\bibfnamefont
  {K.}~\bibnamefont {Latifi}}, \ and\ \bibinfo {author} {\bibfnamefont
  {C.}~\bibnamefont {Thompson}},\ }\href {\doibase 10.1063/1.2364060}
  {\bibfield  {journal} {\bibinfo  {journal} {Applied Physics Letters}\
  }\textbf {\bibinfo {volume} {89}},\ \bibinfo {pages} {161915} (\bibinfo
  {year} {2006})}\BibitemShut {NoStop}%
\bibitem [{\citenamefont {Perret}\ \emph {et~al.}(2014)\citenamefont {Perret},
  \citenamefont {Highland}, \citenamefont {Stephenson}, \citenamefont
  {Streiffer}, \citenamefont {Zapol}, \citenamefont {Fuoss}, \citenamefont
  {Munkholm},\ and\ \citenamefont {Thompson}}]{2014_Perret_APL105_051602}%
  \BibitemOpen
  \bibfield  {author} {\bibinfo {author} {\bibfnamefont {E.}~\bibnamefont
  {Perret}}, \bibinfo {author} {\bibfnamefont {M.~J.}\ \bibnamefont
  {Highland}}, \bibinfo {author} {\bibfnamefont {G.~B.}\ \bibnamefont
  {Stephenson}}, \bibinfo {author} {\bibfnamefont {S.~K.}\ \bibnamefont
  {Streiffer}}, \bibinfo {author} {\bibfnamefont {P.}~\bibnamefont {Zapol}},
  \bibinfo {author} {\bibfnamefont {P.~H.}\ \bibnamefont {Fuoss}}, \bibinfo
  {author} {\bibfnamefont {A.}~\bibnamefont {Munkholm}}, \ and\ \bibinfo
  {author} {\bibfnamefont {C.}~\bibnamefont {Thompson}},\ }\href {\doibase
  10.1063/1.4892349} {\bibfield  {journal} {\bibinfo  {journal} {Applied
  Physics Letters}\ }\textbf {\bibinfo {volume} {105}},\ \bibinfo {eid}
  {051602} (\bibinfo {year} {2014})}\BibitemShut {NoStop}%
\bibitem [{\citenamefont {Takeda}\ \emph {et~al.}(2011)\citenamefont {Takeda},
  \citenamefont {Ninoi}, \citenamefont {Ju}, \citenamefont {Kamiya},
  \citenamefont {Mizuno}, \citenamefont {Fuchi},\ and\ \citenamefont
  {Tabuchi}}]{2011_takeda_2011x}%
  \BibitemOpen
  \bibfield  {author} {\bibinfo {author} {\bibfnamefont {Y.}~\bibnamefont
  {Takeda}}, \bibinfo {author} {\bibfnamefont {K.}~\bibnamefont {Ninoi}},
  \bibinfo {author} {\bibfnamefont {G.}~\bibnamefont {Ju}}, \bibinfo {author}
  {\bibfnamefont {H.}~\bibnamefont {Kamiya}}, \bibinfo {author} {\bibfnamefont
  {T.}~\bibnamefont {Mizuno}}, \bibinfo {author} {\bibfnamefont
  {S.}~\bibnamefont {Fuchi}}, \ and\ \bibinfo {author} {\bibfnamefont
  {M.}~\bibnamefont {Tabuchi}},\ }\href {\doibase
  10.1088/1757-899X/24/1/012002} {\bibfield  {journal} {\bibinfo  {journal}
  {IOP Conference Series: Materials Science and Engineering}\ }\textbf
  {\bibinfo {volume} {24}},\ \bibinfo {pages} {012002} (\bibinfo {year}
  {2011})}\BibitemShut {NoStop}%
\bibitem [{\citenamefont {Iida}\ \emph {et~al.}(2013)\citenamefont {Iida},
  \citenamefont {Kondo}, \citenamefont {Sowa}, \citenamefont {Sugiyama},
  \citenamefont {Iwaya}, \citenamefont {Takeuchi}, \citenamefont {Kamiyama},\
  and\ \citenamefont {Akasaki}}]{iida2013analysis}%
  \BibitemOpen
  \bibfield  {author} {\bibinfo {author} {\bibfnamefont {D.}~\bibnamefont
  {Iida}}, \bibinfo {author} {\bibfnamefont {Y.}~\bibnamefont {Kondo}},
  \bibinfo {author} {\bibfnamefont {M.}~\bibnamefont {Sowa}}, \bibinfo {author}
  {\bibfnamefont {T.}~\bibnamefont {Sugiyama}}, \bibinfo {author}
  {\bibfnamefont {M.}~\bibnamefont {Iwaya}}, \bibinfo {author} {\bibfnamefont
  {T.}~\bibnamefont {Takeuchi}}, \bibinfo {author} {\bibfnamefont
  {S.}~\bibnamefont {Kamiyama}}, \ and\ \bibinfo {author} {\bibfnamefont
  {I.}~\bibnamefont {Akasaki}},\ }\href {\doibase 10.1002/pssr.201307023}
  {\bibfield  {journal} {\bibinfo  {journal} {physica status solidi (RRL)-Rapid
  Research Letters}\ }\textbf {\bibinfo {volume} {7}},\ \bibinfo {pages} {211}
  (\bibinfo {year} {2013})}\BibitemShut {NoStop}%
\bibitem [{\citenamefont {Ju}\ \emph {et~al.}(2014{\natexlab{a}})\citenamefont
  {Ju}, \citenamefont {Fuchi}, \citenamefont {Tabuchi}, \citenamefont {Amano},\
  and\ \citenamefont {Takeda}}]{2014_Ju_JCrystGrowth407_68}%
  \BibitemOpen
  \bibfield  {author} {\bibinfo {author} {\bibfnamefont {G.}~\bibnamefont
  {Ju}}, \bibinfo {author} {\bibfnamefont {S.}~\bibnamefont {Fuchi}}, \bibinfo
  {author} {\bibfnamefont {M.}~\bibnamefont {Tabuchi}}, \bibinfo {author}
  {\bibfnamefont {H.}~\bibnamefont {Amano}}, \ and\ \bibinfo {author}
  {\bibfnamefont {Y.}~\bibnamefont {Takeda}},\ }\href {\doibase
  10.1016/j.jcrysgro.2014.08.023} {\bibfield  {journal} {\bibinfo  {journal}
  {Journal of Crystal Growth}\ }\textbf {\bibinfo {volume} {407}},\ \bibinfo
  {pages} {68} (\bibinfo {year} {2014}{\natexlab{a}})}\BibitemShut {NoStop}%
\bibitem [{\citenamefont {Ju}\ \emph {et~al.}(2014{\natexlab{b}})\citenamefont
  {Ju}, \citenamefont {Honda}, \citenamefont {Tabuchi}, \citenamefont
  {Takeda},\ and\ \citenamefont {Amano}}]{2014_ju_JAP_insitu}%
  \BibitemOpen
  \bibfield  {author} {\bibinfo {author} {\bibfnamefont {G.}~\bibnamefont
  {Ju}}, \bibinfo {author} {\bibfnamefont {Y.}~\bibnamefont {Honda}}, \bibinfo
  {author} {\bibfnamefont {M.}~\bibnamefont {Tabuchi}}, \bibinfo {author}
  {\bibfnamefont {Y.}~\bibnamefont {Takeda}}, \ and\ \bibinfo {author}
  {\bibfnamefont {H.}~\bibnamefont {Amano}},\ }\href {\doibase
  10.1063/1.4867640} {\bibfield  {journal} {\bibinfo  {journal} {Journal of
  Applied Physics}\ }\textbf {\bibinfo {volume} {115}},\ \bibinfo {pages}
  {094906} (\bibinfo {year} {2014}{\natexlab{b}})}\BibitemShut {NoStop}%
\bibitem [{\citenamefont {Headrick}\ \emph {et~al.}(1998)\citenamefont
  {Headrick}, \citenamefont {Kycia}, \citenamefont {Woll}, \citenamefont
  {Brock},\ and\ \citenamefont {Murty}}]{1998_Headrick_PRB58_4818}%
  \BibitemOpen
  \bibfield  {author} {\bibinfo {author} {\bibfnamefont {R.~L.}\ \bibnamefont
  {Headrick}}, \bibinfo {author} {\bibfnamefont {S.}~\bibnamefont {Kycia}},
  \bibinfo {author} {\bibfnamefont {A.~R.}\ \bibnamefont {Woll}}, \bibinfo
  {author} {\bibfnamefont {J.~D.}\ \bibnamefont {Brock}}, \ and\ \bibinfo
  {author} {\bibfnamefont {R.~M.~V.}\ \bibnamefont {Murty}},\ }\href {\doibase
  10.1103/PhysRevB.58.4818} {\bibfield  {journal} {\bibinfo  {journal}
  {Physical Review B}\ }\textbf {\bibinfo {volume} {58}},\ \bibinfo {pages}
  {4818} (\bibinfo {year} {1998})}\BibitemShut {NoStop}%
\bibitem [{\citenamefont {Woll}\ \emph {et~al.}(1999)\citenamefont {Woll},
  \citenamefont {Headrick}, \citenamefont {Kycia},\ and\ \citenamefont
  {Brock}}]{1999_Woll_PRL83_4349}%
  \BibitemOpen
  \bibfield  {author} {\bibinfo {author} {\bibfnamefont {A.}~\bibnamefont
  {Woll}}, \bibinfo {author} {\bibfnamefont {R.}~\bibnamefont {Headrick}},
  \bibinfo {author} {\bibfnamefont {S.}~\bibnamefont {Kycia}}, \ and\ \bibinfo
  {author} {\bibfnamefont {J.}~\bibnamefont {Brock}},\ }\href {\doibase
  10.1103/PhysRevLett.83.4349} {\bibfield  {journal} {\bibinfo  {journal}
  {Physical Review Letters}\ }\textbf {\bibinfo {volume} {83}},\ \bibinfo
  {pages} {4349} (\bibinfo {year} {1999})}\BibitemShut {NoStop}%
\bibitem [{\citenamefont {Sasaki}\ \emph {et~al.}(2016)\citenamefont {Sasaki},
  \citenamefont {Ishikawa}, \citenamefont {Yamaguchi},\ and\ \citenamefont
  {Takahasi}}]{2016_Sasaki_JJAP55_05FB05}%
  \BibitemOpen
  \bibfield  {author} {\bibinfo {author} {\bibfnamefont {T.}~\bibnamefont
  {Sasaki}}, \bibinfo {author} {\bibfnamefont {F.}~\bibnamefont {Ishikawa}},
  \bibinfo {author} {\bibfnamefont {T.}~\bibnamefont {Yamaguchi}}, \ and\
  \bibinfo {author} {\bibfnamefont {M.}~\bibnamefont {Takahasi}},\ }\href
  {\doibase 10.7567/JJAP.55.05FB05} {\bibfield  {journal} {\bibinfo  {journal}
  {Japanese Journal of Applied Physics}\ }\textbf {\bibinfo {volume} {55}},\
  \bibinfo {pages} {05FB05} (\bibinfo {year} {2016})}\BibitemShut {NoStop}%
\bibitem [{\citenamefont {Kang}, \citenamefont {Seo},\ and\ \citenamefont
  {Noh}(2001)}]{2001_Kang_JMaterRes16_1814}%
  \BibitemOpen
  \bibfield  {author} {\bibinfo {author} {\bibfnamefont {H.~C.}\ \bibnamefont
  {Kang}}, \bibinfo {author} {\bibfnamefont {S.~H.}\ \bibnamefont {Seo}}, \
  and\ \bibinfo {author} {\bibfnamefont {D.~Y.}\ \bibnamefont {Noh}},\ }\href
  {\doibase 10.1557/JMR.2001.0250} {\bibfield  {journal} {\bibinfo  {journal}
  {Journal of Materials Research}\ }\textbf {\bibinfo {volume} {16}},\ \bibinfo
  {pages} {1814} (\bibinfo {year} {2001})}\BibitemShut {NoStop}%
\bibitem [{\citenamefont {Fuoss}\ and\ \citenamefont
  {Brennan}(1990)}]{1990_Fuoss_AnnRevMatSci20_365}%
  \BibitemOpen
  \bibfield  {author} {\bibinfo {author} {\bibfnamefont {P.~H.}\ \bibnamefont
  {Fuoss}}\ and\ \bibinfo {author} {\bibfnamefont {S.}~\bibnamefont
  {Brennan}},\ }\href {\doibase 10.1146/annurev.ms.20.080190.002053} {\bibfield
   {journal} {\bibinfo  {journal} {Annual Review of Materials Science}\
  }\textbf {\bibinfo {volume} {20}},\ \bibinfo {pages} {365} (\bibinfo {year}
  {1990})}\BibitemShut {NoStop}%
\bibitem [{\citenamefont {Stephenson}, \citenamefont {Robert},\ and\
  \citenamefont {Gr{\"u}bel}(2009)}]{2009_stephenson_naturematerials}%
  \BibitemOpen
  \bibfield  {author} {\bibinfo {author} {\bibfnamefont {G.~B.}\ \bibnamefont
  {Stephenson}}, \bibinfo {author} {\bibfnamefont {A.}~\bibnamefont {Robert}},
  \ and\ \bibinfo {author} {\bibfnamefont {G.}~\bibnamefont {Gr{\"u}bel}},\
  }\href {\doibase 10.1038/nmat2521} {\bibfield  {journal} {\bibinfo  {journal}
  {Nature Materials}\ }\textbf {\bibinfo {volume} {8}},\ \bibinfo {pages} {702}
  (\bibinfo {year} {2009})}\BibitemShut {NoStop}%
\bibitem [{\citenamefont {Shpyrko}(2014)}]{2014_Shpyrko_JSynchRad21_1057}%
  \BibitemOpen
  \bibfield  {author} {\bibinfo {author} {\bibfnamefont {O.~G.}\ \bibnamefont
  {Shpyrko}},\ }\href {\doibase 10.1107/S1600577514018232} {\bibfield
  {journal} {\bibinfo  {journal} {Journal of Synchrotron Radiation}\ }\textbf
  {\bibinfo {volume} {21}},\ \bibinfo {pages} {1057} (\bibinfo {year}
  {2014})}\BibitemShut {NoStop}%
\bibitem [{\citenamefont {Pierce}\ \emph {et~al.}(2009)\citenamefont {Pierce},
  \citenamefont {Chang}, \citenamefont {Hennessy}, \citenamefont {Komanicky},
  \citenamefont {Sprung}, \citenamefont {Sandy},\ and\ \citenamefont
  {You}}]{pierce2009surface}%
  \BibitemOpen
  \bibfield  {author} {\bibinfo {author} {\bibfnamefont {M.~S.}\ \bibnamefont
  {Pierce}}, \bibinfo {author} {\bibfnamefont {K.~C.}\ \bibnamefont {Chang}},
  \bibinfo {author} {\bibfnamefont {D.}~\bibnamefont {Hennessy}}, \bibinfo
  {author} {\bibfnamefont {V.}~\bibnamefont {Komanicky}}, \bibinfo {author}
  {\bibfnamefont {M.}~\bibnamefont {Sprung}}, \bibinfo {author} {\bibfnamefont
  {A.}~\bibnamefont {Sandy}}, \ and\ \bibinfo {author} {\bibfnamefont {H.-Y.}\
  \bibnamefont {You}},\ }\href {\doibase 10.1103/PhysRevLett.103.165501}
  {\bibfield  {journal} {\bibinfo  {journal} {Physical Review Letters}\
  }\textbf {\bibinfo {volume} {103}},\ \bibinfo {pages} {165501} (\bibinfo
  {year} {2009})}\BibitemShut {NoStop}%
\bibitem [{\citenamefont {Abbey}(2013)}]{2013_Abbey_JOM65_1183}%
  \BibitemOpen
  \bibfield  {author} {\bibinfo {author} {\bibfnamefont {B.}~\bibnamefont
  {Abbey}},\ }\href {\doibase 10.1007/s11837-013-0702-4} {\bibfield  {journal}
  {\bibinfo  {journal} {{JOM}}\ }\textbf {\bibinfo {volume} {65}},\ \bibinfo
  {pages} {1183} (\bibinfo {year} {2013})}\BibitemShut {NoStop}%
\bibitem [{\citenamefont {Zhu}\ \emph {et~al.}(2015)\citenamefont {Zhu},
  \citenamefont {Harder}, \citenamefont {Diaz}, \citenamefont {Komanicky},
  \citenamefont {Barbour}, \citenamefont {Xu}, \citenamefont {Huang},
  \citenamefont {Liu}, \citenamefont {Pierce}, \citenamefont {Menzel} \emph
  {et~al.}}]{2015_Zhu_APL106_101604}%
  \BibitemOpen
  \bibfield  {author} {\bibinfo {author} {\bibfnamefont {C.}~\bibnamefont
  {Zhu}}, \bibinfo {author} {\bibfnamefont {R.}~\bibnamefont {Harder}},
  \bibinfo {author} {\bibfnamefont {A.}~\bibnamefont {Diaz}}, \bibinfo {author}
  {\bibfnamefont {V.}~\bibnamefont {Komanicky}}, \bibinfo {author}
  {\bibfnamefont {A.}~\bibnamefont {Barbour}}, \bibinfo {author} {\bibfnamefont
  {R.}~\bibnamefont {Xu}}, \bibinfo {author} {\bibfnamefont {X.}~\bibnamefont
  {Huang}}, \bibinfo {author} {\bibfnamefont {Y.}~\bibnamefont {Liu}}, \bibinfo
  {author} {\bibfnamefont {M.~S.}\ \bibnamefont {Pierce}}, \bibinfo {author}
  {\bibfnamefont {A.}~\bibnamefont {Menzel}},  \emph {et~al.},\ }\href
  {\doibase 10.1063/1.4914927} {\bibfield  {journal} {\bibinfo  {journal}
  {Applied Physics Letters}\ }\textbf {\bibinfo {volume} {106}},\ \bibinfo
  {pages} {101604} (\bibinfo {year} {2015})}\BibitemShut {NoStop}%
\bibitem [{\citenamefont {Hruszkewycz}\ \emph {et~al.}(2012)\citenamefont
  {Hruszkewycz}, \citenamefont {Holt}, \citenamefont {Murray}, \citenamefont
  {Bruley}, \citenamefont {Holt}, \citenamefont {Tripathi}, \citenamefont
  {Shpyrko}, \citenamefont {McNulty}, \citenamefont {Highland},\ and\
  \citenamefont {Fuoss}}]{hruszkewycz2012quantitative}%
  \BibitemOpen
  \bibfield  {author} {\bibinfo {author} {\bibfnamefont {S.~O.}\ \bibnamefont
  {Hruszkewycz}}, \bibinfo {author} {\bibfnamefont {M.~V.}\ \bibnamefont
  {Holt}}, \bibinfo {author} {\bibfnamefont {C.~E.}\ \bibnamefont {Murray}},
  \bibinfo {author} {\bibfnamefont {J.}~\bibnamefont {Bruley}}, \bibinfo
  {author} {\bibfnamefont {J.}~\bibnamefont {Holt}}, \bibinfo {author}
  {\bibfnamefont {A.}~\bibnamefont {Tripathi}}, \bibinfo {author}
  {\bibfnamefont {O.~G.}\ \bibnamefont {Shpyrko}}, \bibinfo {author}
  {\bibfnamefont {I.}~\bibnamefont {McNulty}}, \bibinfo {author} {\bibfnamefont
  {M.~J.}\ \bibnamefont {Highland}}, \ and\ \bibinfo {author} {\bibfnamefont
  {P.~H.}\ \bibnamefont {Fuoss}},\ }\href {\doibase 10.1021/nl303201w}
  {\bibfield  {journal} {\bibinfo  {journal} {Nano Letters}\ }\textbf {\bibinfo
  {volume} {12}},\ \bibinfo {pages} {5148} (\bibinfo {year}
  {2012})}\BibitemShut {NoStop}%
\bibitem [{201()}]{2015_APS_Early_Science}%
  \BibitemOpen
  \href@noop {} {}\bibinfo {note} {Early Science at the Upgraded Advanced
  Photon Source, (2015),
  \url{https://www1.aps.anl.gov/files/download/Aps-Upgrade/Beamlines/APS-U
  Early-Science-103015-FINAL.pdf}}\BibitemShut {NoStop}%
\bibitem [{\citenamefont {Dejus}(2016)}]{Dejus_private}%
  \BibitemOpen
  \bibfield  {author} {\bibinfo {author} {\bibfnamefont {R.}~\bibnamefont
  {Dejus}},\ }\href@noop {} {} (\bibinfo {year} {2016}),\ \bibinfo {note}
  {private communication}\BibitemShut {NoStop}%
\bibitem [{\citenamefont {Nakamura}, \citenamefont {Harada},\ and\
  \citenamefont {Seno}(1991)}]{nakamura1991novel}%
  \BibitemOpen
  \bibfield  {author} {\bibinfo {author} {\bibfnamefont {S.}~\bibnamefont
  {Nakamura}}, \bibinfo {author} {\bibfnamefont {Y.}~\bibnamefont {Harada}}, \
  and\ \bibinfo {author} {\bibfnamefont {M.}~\bibnamefont {Seno}},\ }\href
  {\doibase 10.1063/1.105239} {\bibfield  {journal} {\bibinfo  {journal}
  {Applied Physics Letters}\ }\textbf {\bibinfo {volume} {58}},\ \bibinfo
  {pages} {2021} (\bibinfo {year} {1991})}\BibitemShut {NoStop}%
\bibitem [{\citenamefont {Brennan}\ \emph {et~al.}(1990)\citenamefont
  {Brennan}, \citenamefont {Fuoss}, \citenamefont {Kahn},\ and\ \citenamefont
  {Kisker}}]{1990_Brennan_NuclInstMeth291_86}%
  \BibitemOpen
  \bibfield  {author} {\bibinfo {author} {\bibfnamefont {S.}~\bibnamefont
  {Brennan}}, \bibinfo {author} {\bibfnamefont {P.~H.}\ \bibnamefont {Fuoss}},
  \bibinfo {author} {\bibfnamefont {J.~L.}\ \bibnamefont {Kahn}}, \ and\
  \bibinfo {author} {\bibfnamefont {D.~W.}\ \bibnamefont {Kisker}},\ }\href
  {\doibase 10.1016/0168-9002(90)90038-8} {\bibfield  {journal} {\bibinfo
  {journal} {Nuclear Instruments \& Methods In Physics Research Section
  A-Accelerators Spectrometers Detectors and Associated Equipment}\ }\textbf
  {\bibinfo {volume} {291}},\ \bibinfo {pages} {86} (\bibinfo {year}
  {1990})}\BibitemShut {NoStop}%
\bibitem [{The()}]{Thermionics}%
  \BibitemOpen
  \href@noop {} {}\bibinfo {note} {Thermionics, RNN Series Differentially
  Pumped Rotary Seals, \url{http://www.thermionics.com}}\BibitemShut {NoStop}%
\bibitem [{Bor()}]{Boralectric}%
  \BibitemOpen
  \href@noop {} {}\bibinfo {note}
  {Boralectric\textsuperscript{\textregistered}, GE Advanced Ceramics Corp or
  Momentive Performance Materials}\BibitemShut {NoStop}%
\bibitem [{\citenamefont {Laanait}\ \emph {et~al.}(2014)\citenamefont
  {Laanait}, \citenamefont {Zhang}, \citenamefont {Schlep{\"u}tz},
  \citenamefont {Vila-Comamala}, \citenamefont {Highland},\ and\ \citenamefont
  {Fenter}}]{2014_Laanait_JSynchRad21_1252}%
  \BibitemOpen
  \bibfield  {author} {\bibinfo {author} {\bibfnamefont {N.}~\bibnamefont
  {Laanait}}, \bibinfo {author} {\bibfnamefont {Z.}~\bibnamefont {Zhang}},
  \bibinfo {author} {\bibfnamefont {C.~M.}\ \bibnamefont {Schlep{\"u}tz}},
  \bibinfo {author} {\bibfnamefont {J.}~\bibnamefont {Vila-Comamala}}, \bibinfo
  {author} {\bibfnamefont {M.~J.}\ \bibnamefont {Highland}}, \ and\ \bibinfo
  {author} {\bibfnamefont {P.}~\bibnamefont {Fenter}},\ }\href {\doibase
  10.1107/S1600577514016555} {\bibfield  {journal} {\bibinfo  {journal}
  {Journal of Synchrotron Radiation}\ }\textbf {\bibinfo {volume} {21}},\
  \bibinfo {pages} {1252} (\bibinfo {year} {2014})}\BibitemShut {NoStop}%
\bibitem [{Hub()}]{Huber}%
  \BibitemOpen
  \href@noop {} {}\bibinfo {note} {HUBER Diffraktionstechnik GmbH \& Co.,
  Rimsting, Germany, \url{http://www.xhuber.de/en/}}\BibitemShut {NoStop}%
\bibitem [{Sym()}]{Symmetrie}%
  \BibitemOpen
  \href@noop {} {}\bibinfo {note} {Breva Hexapod, Symm{\'e}trie, 10, All{\'e}e
  Charles Babbage 30035 NIMES Cedex 1, France,
  \url{http://www.symetrie.fr/en/positioning/products/breva-hexapod/}}\BibitemShut
  {NoStop}%
\bibitem [{EPI()}]{EPICSwebsite}%
  \BibitemOpen
  \href@noop {} {}\bibinfo {note} {Description of the open source Experimental
  Physics and Industrial Control System (EPICS) can be found at
  \url{http://www.aps.anl.gov/epics/}}\BibitemShut {NoStop}%
\bibitem [{SPE()}]{SPECwebsite}%
  \BibitemOpen
  \href@noop {} {}\bibinfo {note} {Certified Scientific Software, Cambridge,
  MA, USA, \url{https://certif.com/spec_manual/idx.html}}\BibitemShut {NoStop}%
\bibitem [{\citenamefont {Kee}, \citenamefont {Coltrin},\ and\ \citenamefont
  {Glarborg}(2005)}]{kee2005chemically}%
  \BibitemOpen
  \bibfield  {author} {\bibinfo {author} {\bibfnamefont {R.~J.}\ \bibnamefont
  {Kee}}, \bibinfo {author} {\bibfnamefont {M.~E.}\ \bibnamefont {Coltrin}}, \
  and\ \bibinfo {author} {\bibfnamefont {P.}~\bibnamefont {Glarborg}},\ }\href
  {\doibase 10.1002/0471461296} {\emph {\bibinfo {title} {Chemically reacting
  flow: theory and practice}}}\ (\bibinfo  {publisher} {John Wiley \& Sons},\
  \bibinfo {year} {2005})\BibitemShut {NoStop}%
\bibitem [{\citenamefont {Ferziger}\ and\ \citenamefont
  {Peric}(1999)}]{Ferzinger_cfd_1999}%
  \BibitemOpen
  \bibfield  {author} {\bibinfo {author} {\bibfnamefont {J.}~\bibnamefont
  {Ferziger}}\ and\ \bibinfo {author} {\bibfnamefont {M.}~\bibnamefont
  {Peric}},\ }\href@noop {} {\emph {\bibinfo {title} {Computational methods for
  fluid dynamics}}}\ (\bibinfo  {publisher} {Springer},\ \bibinfo {address}
  {Berlin},\ \bibinfo {year} {1999})\BibitemShut {NoStop}%
\bibitem [{\citenamefont {Weller}\ \emph {et~al.}(1998)\citenamefont {Weller},
  \citenamefont {Tabor}, \citenamefont {Jasak},\ and\ \citenamefont
  {Fureby}}]{weller1998tensorial}%
  \BibitemOpen
  \bibfield  {author} {\bibinfo {author} {\bibfnamefont {H.~G.}\ \bibnamefont
  {Weller}}, \bibinfo {author} {\bibfnamefont {G.}~\bibnamefont {Tabor}},
  \bibinfo {author} {\bibfnamefont {H.}~\bibnamefont {Jasak}}, \ and\ \bibinfo
  {author} {\bibfnamefont {C.}~\bibnamefont {Fureby}},\ }\href {\doibase
  10.1063/1.168744} {\bibfield  {journal} {\bibinfo  {journal} {Computers in
  Physics}\ }\textbf {\bibinfo {volume} {12}},\ \bibinfo {pages} {620}
  (\bibinfo {year} {1998})}\BibitemShut {NoStop}%
\bibitem [{\citenamefont {Saenger}\ and\ \citenamefont
  {Gupta}(1991)}]{1991_Saenger_ApplOpt30_1221}%
  \BibitemOpen
  \bibfield  {author} {\bibinfo {author} {\bibfnamefont {K.~L.}\ \bibnamefont
  {Saenger}}\ and\ \bibinfo {author} {\bibfnamefont {J.}~\bibnamefont
  {Gupta}},\ }\href {\doibase 10.1364/AO.30.001221} {\bibfield  {journal}
  {\bibinfo  {journal} {Applied Optics}\ }\textbf {\bibinfo {volume} {30}},\
  \bibinfo {pages} {1221} (\bibinfo {year} {1991})}\BibitemShut {NoStop}%
\bibitem [{\citenamefont {Tapping}\ and\ \citenamefont
  {Reilly}(1986)}]{1986_Tapping_JOSAA3_610}%
  \BibitemOpen
  \bibfield  {author} {\bibinfo {author} {\bibfnamefont {J.}~\bibnamefont
  {Tapping}}\ and\ \bibinfo {author} {\bibfnamefont {M.~L.}\ \bibnamefont
  {Reilly}},\ }\href {\doibase 10.1364/JOSAA.3.000610} {\bibfield  {journal}
  {\bibinfo  {journal} {Journal of the Optical Society of America A, Optics and
  Image Science}\ }\textbf {\bibinfo {volume} {3}},\ \bibinfo {pages} {610}
  (\bibinfo {year} {1986})}\BibitemShut {NoStop}%
\bibitem [{\citenamefont {Touloulian}\ \emph {et~al.}(1977)\citenamefont
  {Touloulian}, \citenamefont {Kirby}, \citenamefont {Taylor},\ and\
  \citenamefont {Lee}}]{1977_Touloulian_TPRCseries13}%
  \BibitemOpen
  \bibfield  {author} {\bibinfo {author} {\bibfnamefont {Y.~S.}\ \bibnamefont
  {Touloulian}}, \bibinfo {author} {\bibfnamefont {R.~K.}\ \bibnamefont
  {Kirby}}, \bibinfo {author} {\bibfnamefont {R.~E.}\ \bibnamefont {Taylor}}, \
  and\ \bibinfo {author} {\bibfnamefont {T.~Y.}\ \bibnamefont {Lee}},\ }\href
  {http://www.springer.com/series/10794} {\emph {\bibinfo {title} {Thermal
  Expansion: Nonmetallic Solids}}},\ \bibinfo {series} {The TPRC Data Series},
  Vol.~\bibinfo {volume} {13}\ (\bibinfo  {publisher} {Springer},\ \bibinfo
  {year} {1977})\ pp.\ \bibinfo {pages} {154--389}\BibitemShut {NoStop}%
\bibitem [{\citenamefont
  {Hofmeister}(2014)}]{2014_Hofmeister_PhysChemMin41_361}%
  \BibitemOpen
  \bibfield  {author} {\bibinfo {author} {\bibfnamefont {A.~M.}\ \bibnamefont
  {Hofmeister}},\ }\href {\doibase 10.1007/s00269-014-0655-3} {\bibfield
  {journal} {\bibinfo  {journal} {Physics and Chemistry of Minerals}\ }\textbf
  {\bibinfo {volume} {41}},\ \bibinfo {pages} {361} (\bibinfo {year}
  {2014})}\BibitemShut {NoStop}%
\bibitem [{\citenamefont {Watanabe}, \citenamefont {Kimoto},\ and\
  \citenamefont {Suda}(2008)}]{watanabe2008temperature}%
  \BibitemOpen
  \bibfield  {author} {\bibinfo {author} {\bibfnamefont {N.}~\bibnamefont
  {Watanabe}}, \bibinfo {author} {\bibfnamefont {T.}~\bibnamefont {Kimoto}}, \
  and\ \bibinfo {author} {\bibfnamefont {J.}~\bibnamefont {Suda}},\ }\href
  {\doibase 10.1063/1.3021148} {\bibfield  {journal} {\bibinfo  {journal}
  {Journal of Applied Physics}\ }\textbf {\bibinfo {volume} {104}},\ \bibinfo
  {pages} {106101} (\bibinfo {year} {2008})}\BibitemShut {NoStop}%
\bibitem [{\citenamefont {Liu}\ \emph {et~al.}(2006)\citenamefont {Liu},
  \citenamefont {Stepanov}, \citenamefont {Gott}, \citenamefont {Shields},
  \citenamefont {Zhirnov}, \citenamefont {Wang}, \citenamefont {Steimetz},\
  and\ \citenamefont {Zettler}}]{2006_Liu_physstatsolc3_1884}%
  \BibitemOpen
  \bibfield  {author} {\bibinfo {author} {\bibfnamefont {C.}~\bibnamefont
  {Liu}}, \bibinfo {author} {\bibfnamefont {S.}~\bibnamefont {Stepanov}},
  \bibinfo {author} {\bibfnamefont {A.}~\bibnamefont {Gott}}, \bibinfo {author}
  {\bibfnamefont {P.~A.}\ \bibnamefont {Shields}}, \bibinfo {author}
  {\bibfnamefont {E.}~\bibnamefont {Zhirnov}}, \bibinfo {author} {\bibfnamefont
  {W.~N.}\ \bibnamefont {Wang}}, \bibinfo {author} {\bibfnamefont
  {E.}~\bibnamefont {Steimetz}}, \ and\ \bibinfo {author} {\bibfnamefont
  {J.~T.}\ \bibnamefont {Zettler}},\ }\href {\doibase 10.1002/pssc.200565197}
  {\bibfield  {journal} {\bibinfo  {journal} {physica status solidi (c)}\
  }\textbf {\bibinfo {volume} {3}},\ \bibinfo {pages} {1884} (\bibinfo {year}
  {2006})}\BibitemShut {NoStop}%
\bibitem [{\citenamefont {Reeber}\ and\ \citenamefont
  {Wang}(2000)}]{2000_Reeber_JMR15_40}%
  \BibitemOpen
  \bibfield  {author} {\bibinfo {author} {\bibfnamefont {R.~R.}\ \bibnamefont
  {Reeber}}\ and\ \bibinfo {author} {\bibfnamefont {K.}~\bibnamefont {Wang}},\
  }\href {\doibase 10.1557/JMR.2000.0011} {\bibfield  {journal} {\bibinfo
  {journal} {Journal of Materials Research}\ }\textbf {\bibinfo {volume}
  {15}},\ \bibinfo {pages} {40} (\bibinfo {year} {2000})}\BibitemShut {NoStop}%
\bibitem [{\citenamefont {Shibata}\ \emph {et~al.}(2007)\citenamefont
  {Shibata}, \citenamefont {Waseda}, \citenamefont {Ohta}, \citenamefont
  {Kiyomi}, \citenamefont {Shimoyama}, \citenamefont {Fujito}, \citenamefont
  {Nagaoka}, \citenamefont {Kagamitani}, \citenamefont {Simura},\ and\
  \citenamefont {Fukuda}}]{2007_Shibata_MaterialsTrans48_2782}%
  \BibitemOpen
  \bibfield  {author} {\bibinfo {author} {\bibfnamefont {H.}~\bibnamefont
  {Shibata}}, \bibinfo {author} {\bibfnamefont {Y.}~\bibnamefont {Waseda}},
  \bibinfo {author} {\bibfnamefont {H.}~\bibnamefont {Ohta}}, \bibinfo {author}
  {\bibfnamefont {K.}~\bibnamefont {Kiyomi}}, \bibinfo {author} {\bibfnamefont
  {K.}~\bibnamefont {Shimoyama}}, \bibinfo {author} {\bibfnamefont
  {K.}~\bibnamefont {Fujito}}, \bibinfo {author} {\bibfnamefont
  {H.}~\bibnamefont {Nagaoka}}, \bibinfo {author} {\bibfnamefont
  {Y.}~\bibnamefont {Kagamitani}}, \bibinfo {author} {\bibfnamefont
  {R.}~\bibnamefont {Simura}}, \ and\ \bibinfo {author} {\bibfnamefont
  {T.}~\bibnamefont {Fukuda}},\ }\href {\doibase 10.2320/matertrans.MRP2007109}
  {\bibfield  {journal} {\bibinfo  {journal} {MATERIALS TRANSACTIONS}\ }\textbf
  {\bibinfo {volume} {48}},\ \bibinfo {pages} {2782} (\bibinfo {year}
  {2007})}\BibitemShut {NoStop}%
\bibitem [{\citenamefont {Chase}(1998)}]{1998_Chase_JPCRDMono9}%
  \BibitemOpen
  \bibfield  {author} {\bibinfo {author} {\bibfnamefont {M.~W.}\ \bibnamefont
  {Chase}, \bibfnamefont {Jr.}},\ }\href
  {https://www.nist.gov/srd/journal-physical-and-chemical-reference-data-monographs-or-supplements}
  {\emph {\bibinfo {title} {NIST-JANAF Thermochemical Tables}}},\ \bibinfo
  {edition} {4th}\ ed.,\ \bibinfo {series} {Journal of Physical and Chemical
  Reference Data: Monograph}\ No.\ \bibinfo {number} {9 (Part I and Part II)}\
  (\bibinfo  {publisher} {American Chemical Society and American Institute of
  Physics for the National Institute of Standards and Technology},\ \bibinfo
  {year} {1998})\BibitemShut {NoStop}%
\bibitem [{\citenamefont {Sutton}\ \emph {et~al.}(1991)\citenamefont {Sutton},
  \citenamefont {Mochrie}, \citenamefont {Greytak}, \citenamefont {Nagler},
  \citenamefont {Berman}, \citenamefont {Held},\ and\ \citenamefont
  {Stephenson}}]{sutton1991observation}%
  \BibitemOpen
  \bibfield  {author} {\bibinfo {author} {\bibfnamefont {M.}~\bibnamefont
  {Sutton}}, \bibinfo {author} {\bibfnamefont {S.~G.~J.}\ \bibnamefont
  {Mochrie}}, \bibinfo {author} {\bibfnamefont {T.}~\bibnamefont {Greytak}},
  \bibinfo {author} {\bibfnamefont {S.~E.}\ \bibnamefont {Nagler}}, \bibinfo
  {author} {\bibfnamefont {L.~E.}\ \bibnamefont {Berman}}, \bibinfo {author}
  {\bibfnamefont {G.~A.}\ \bibnamefont {Held}}, \ and\ \bibinfo {author}
  {\bibfnamefont {G.~B.}\ \bibnamefont {Stephenson}},\ }\href {\doibase
  10.1038/352608a0} {\bibfield  {journal} {\bibinfo  {journal} {Nature}\
  }\textbf {\bibinfo {volume} {352}},\ \bibinfo {pages} {608} (\bibinfo {year}
  {1991})}\BibitemShut {NoStop}%
\bibitem [{\citenamefont {Malik}\ \emph {et~al.}(1998)\citenamefont {Malik},
  \citenamefont {Sandy}, \citenamefont {Lurio}, \citenamefont {Stephenson},
  \citenamefont {Mochrie}, \citenamefont {McNulty},\ and\ \citenamefont
  {Sutton}}]{malik1998coherent}%
  \BibitemOpen
  \bibfield  {author} {\bibinfo {author} {\bibfnamefont {A.}~\bibnamefont
  {Malik}}, \bibinfo {author} {\bibfnamefont {A.~R.}\ \bibnamefont {Sandy}},
  \bibinfo {author} {\bibfnamefont {L.~B.}\ \bibnamefont {Lurio}}, \bibinfo
  {author} {\bibfnamefont {G.~B.}\ \bibnamefont {Stephenson}}, \bibinfo
  {author} {\bibfnamefont {S.~G.~J.}\ \bibnamefont {Mochrie}}, \bibinfo
  {author} {\bibfnamefont {I.}~\bibnamefont {McNulty}}, \ and\ \bibinfo
  {author} {\bibfnamefont {M.}~\bibnamefont {Sutton}},\ }\href {\doibase
  10.1103/PhysRevLett.81.5832} {\bibfield  {journal} {\bibinfo  {journal}
  {Physical Review Letters}\ }\textbf {\bibinfo {volume} {81}},\ \bibinfo
  {pages} {5832} (\bibinfo {year} {1998})}\BibitemShut {NoStop}%
\end{thebibliography}
